\documentclass[aps,pra,twocolumn,showpacs,groupedaddress]{revtex4}
\usepackage{graphicx}
\def\bra{\langle}
\def\ket{\rangle}

\def\su{\uparrow}
\def\sd{\downarrow}
\def\bs{\bar{s}}
\begin{document}

\title{Boson-Fermion coherence in a spherically symmetric harmonic trap}
\author{Takahiko Miyakawa and Pierre Meystre}
\affiliation{Optical Sciences Center,
The University of Arizona, Tucson, AZ 85721}

\date{\today}

\begin{abstract}
We consider the photoassociation of a low-density gas of
quantum-degenerate trapped fermionic atoms into bosonic molecules
in a spherically symmetric harmonic potential. For a dilute system
and the photoassociation coupling energy small
compared to the level separation of the trap, only those fermions
in the single shell with Fermi energy are coupled to the bosonic
molecular field.
Introducing a collective pseudo-spin operator formalism we show
that this system can then be mapped onto the Tavis-Cummings
Hamiltonian of quantum optics, with an additional pairing
interaction. By exact diagonalization of the Hamiltonian, we
examine the ground state and low excitations of the Bose-Fermi
system, and study the dynamics of the coherent coupling between
atoms and molecules. In a semiclassical description of the system,
the pairing interaction between fermions is shown to result in a
self-trapping transition in the photoassociation, with a sudden
suppression of the coherent oscillations between atoms and
molecules. We also show that the full quantum dynamics of the
system is dominated by quantum fluctuations in the vicinity of the
self-trapping solution.
\end{abstract}

 \maketitle
\section{Introduction}

The formation of ultracold diatomic molecules from Feshbach
resonances and photoassociation has witnessed spectacular
developments in recent years. Early demonstrations of molecule
formation using two-photon Raman photoassociation \cite{Wynar00}
and a Feshbach resonance \cite{Inouye98} were dominated by the
molecular losses due to processes such as inelastic decay to lower
energy molecular vibrational states \cite{Yurovsky99}, so that the
existence of the molecules could only be inferred from the
decrease in the number of atoms. The first unambiguous coherent
conversion of atoms into molecules was performed by Donley {\it et
al.} \cite{Donley02}, who exploited a Feshbach resonance in a
$^{85}$Rb Bose-Einstein Condensate (BEC). In subsequent experiments starting from an
atomic condensate of $^{87}$Rb, Rempe and coworkers used adiabatic
rapid passage to create the molecules \cite{Durr04}. Because the
molecules and atoms have different magnetic moments, they could be
spatially separated from each other using a magnetic field
gradient via the Stern-Gerlach effect. Similar work has been
conducted by Xu {\it et al.} \cite{Xu03}. Starting from a sodium
BEC, they used resonant laser light to blast away the remaining
atoms in the sample and isolated the molecules. Unfortunately, the
conversion efficiency was limited by inelastic losses very close
to the resonance so that molecular yields were $\lesssim 10\%$.

For fermionic atoms close to a Feshbach resonance the inelastic
collision rate for relaxation  to lower energy vibrational states
of the molecules scales like $a(B)^{-2.55}$ whereas for bosons it
scales like $a(B)$ where $a(B)$ is the scattering length near the
resonance \cite{Petrov04}. This is because close to resonance the
effective size of the molecules is of the order $a(B)$, which is
comparable to the interparticle spacing. In order for a molecule
to decay to a more deeply bound vibrational state with radius
$R_e\ll a(B)$, the atoms comprising the molecule along with an
additional atom must all collide within a distance $\sim R_e$.
Since two of the three atoms are necessarily identical, the
collision rate is suppressed for fermions. Taking advantage of
this consequence of the Pauli Exclusion Principle, Greiner {\it et
al.} \cite{Greiner03} achieved the first molecular BEC by starting from a quantum degenerate spin mixture of
$^{40}$K using adiabatic rapid passage through a Feshbach
resonance with a conversion efficiency of $\sim 80\%$. At about
the same time Zwierlein {\it et al.} \cite{Zwierlein03} and Jochim
{\it et al.} \cite{Jochim03} succeeded in producing a BEC of
$^{6}$Li$_2$ dimers by evaporatively cooling the atoms at a
constant magnetic field just below a resonance where $a(B)$ is
large and positive. Two recent experiments have led to the
observation of heteronuclear Feshbach resonances in Bose-Fermi
mixtures of $^6$Li and $^{23}$Na \cite{Stan04} in one case, and of
$^{87}$Rb and $^{40}$K in the other \cite{Inouye04}. We also
mention recent experiments by Kerman {\it et al.} \cite{Kerman04},
who produced metastable RbCs molecules in their lowest triplet
state starting from a laser-cooled mixture of $^{85}$Rb and
$^{133}$Cs by photoassociation. Major current experimental and
experimental efforts are directed towards the exploration of the
crossover between the Bardeen-Cooper-Schrieffer (BCS) state of
fermionic atoms and BEC of bosonic
molecules as the system crosses a Feshbach resonance
\cite{ex_BECBCS}. In the vicinity of these resonances, the system
enters a strongly interacting regime that offers a challenge for
many-body theories \cite{Timmermans,th_BECBCS}.

It is known that the pairing properties of finite-size systems can
be significantly different from those of the bulk material, due to
the discrete energy spectrum of the particles involved. The
detailed role of the shell structure has been explored both in the
nuclei~\cite{Mottelson}  and of superconductor
grains~\cite{SCgrain}, in which a collective character of pairs
plays an important role. In this paper we consider the
photoassociation of a dilute quantum-degenerate gas of fermionic
atoms trapped in a spherically symmetric harmonic trap into
molecular dimers, including the pairing interaction between
fermions. By introducing a pseudo-spin formalism for the time
reversal pairing operator~\cite{Anderson,Kerman61}, this problem
can be mapped onto an extension of the Tavis-Cummings
model~\cite{Tavis,Bogoliubov} that describes the coupling of $N$
two-level atoms and a single mode of the electromagnetic field,
and has found applications in the study of superradiance in
quantum optics~\cite{Dicke,Bonifacio}. This analogy allows us to
study in detail the ground state and lower excited states of the
system, as well as the coherent dynamics of atom-molecule coupling
in the trap. We show that for appropriate conditions, only those
fermions on the last energy shell of the trap participate in the
photoassociation process, and discuss the impact of the filling of
that shell on molecule formation. We find that depending on the
detuning $\delta$ of the photoassociation laser from the energy
difference between the molecules and atom pairs, the nature of the
ground state changes from being predominantly atomic to
predominantly molecular in nature. We study the crossover between
these two regions in detail, and quantify its property via the
joint coherence of the atomic and molecular fields and the
entanglement entropy of the system.

The coherent conversion of fermions into bosons has been studied
in the homogeneous case by several
authors~\cite{Timmermans,dy_BECBCS,Vardi}. Trapped systems, in
addition to being closer to the experimental situation, present
several unique characteristics: First, the discreteness of the
energy levels eliminates many of the difficulties associated with
a continuum. In addition, the high degeneracy of spherically
symmetric harmonic potentials simplifies significantly the study
of coherent quantum dynamics. Indeed, the problem resembles then
the dynamics of a bosonic Josephson Junction~\cite{BJJ1,BJJ2},
although the nonlinear coupling between fermionic atoms and
bosonic molecules leads to considerably richer dynamics. Moreover
the additional pairing interaction between fermions is shown to
result in a self-trapping transition~\cite{BJJ1,Scott}, with a
sudden suppression of the coherent oscillations.

This paper is organized as follows. Section II discusses our model
and formulates it in terms of a pseudo-spin formalism. Section III
presents results of the static problem where the many-body states
are classified by number of unpaired fermions known as {\it
seniority}~\cite{Racah} in nuclear physics. We show that for a
fixed number of atoms, the ground state of the system is always
the state of minimum seniority. We examine the crossover behavior
of the ground state as the detuning parameter $\delta$ is varied,
as a function of the ratio of the total number of fermionic pairs
and molecules to the degenerate number of the Fermi level. The
entanglement between fermions and bosons is evaluated and found to
be reduced as the pairing interaction becomes stronger. In section
IV, we analyze the coherent dynamics of the nonlinear
atom-molecule coupling. Using a semiclassical factorization
ansatz, we show the appearance of a self-trapping transition in
the presence of pairing interaction. An exact quantum solution
shows that around that transition point the dynamics is
characterized by large quantum fluctuations. Finally, section V is
a summary and outlook. Calculational details are relegated to an
appendix.

\section{Model}

We consider a trapped dilute gas of two-component fermionic atoms
in hyperfine states of spin $\sigma=\su,\sd$ at zero temperature
and coupled to a single-mode gas of bosonic molecules via a
two-photon Raman transition. The trap is assumed to be harmonic
and spherically symmetric, described by the potential
    \begin{equation}
    V_f=\frac{1}{2}m_f \omega_{\rm ho}^2 {\bf r}^2
    \end{equation}
for the atoms, and similarly with $f \rightarrow b$ for the
bosonic molecules.

In the absence of interactions between particles, the trap levels
have the energies
    \begin{equation}
    {\cal E}_{n}=\left(n+\frac{3}{2}\right)\hbar\omega_{\rm
    ho}.
    \end{equation}
where the principal quantum number $n$ is positive or zero. In
order to deal with the high degree of degeneracy of this
potential, it is convenient to introduce the (integer) radial and
angular quantum numbers $n_r$ and $l$, which are positive or zero,
with \cite{CT}
    \begin{equation}
    n= 2n_{r} + l.
    \end{equation}
Each pair $(n_r, l)$ corresponds to a radial wave function
$R_{n,l}(r)$ and hence $(2l+1)$ common eigenfunctions of $V_f(r)$,
${\bf L}^2$ and $L_z$,
    \begin{equation}
    \phi_{n,l,m}({\bf r})= R_{n,l}(r)
    Y_{lm}(\theta,\phi).
    \end{equation}
Taking into account the magnetic quantum number
    \begin{equation}
    -l \le m \le +l,
    \end{equation}
the degeneracy of each level ${\cal E}_n$ is therefore
    \begin{equation}
    \Omega_n = \frac{1}{2} (n + 1)(n +2),
    \end{equation}
and the total number of states up to the shell $n_F$
corresponding to the Fermi energy
    $$
    {\cal E}_F = \left(n_F + \frac{3}{2}\right) \hbar \omega_{\rm ho}
    $$
is
\[
    N_{n_F}=2\times
    \sum_{n=0}^{n_F}\Omega_{n}=\frac{1}{3}(n_F+1)
    (n_F+2)(n_F+3),
\]
where the factor of 2 accounts for the two hyperfine spin states
of the atoms. In the following, it will be necessary to include
the attractive interaction responsible for the pairing between
fermions. Since as we show later on this interaction splits the
degeneracy of the trap levels, we keep the angular momentum index
in the labelling of the atomic energies, ${\cal E}_{n,l}$. The
fermionic atoms are therefore described by the annihilation
operator $c_{n;l,m,\sigma}$, where $\sigma$ labels the hyperfine
spin state of the atom with single-particle wave function
$\phi_{n,l,m}({\bf r})$ and eigenenergy ${\cal E}_{n,l}$.

Assuming that the atom-molecule photoassociation energy is smaller
than the trap energy spacing, $\hbar\omega_{\rm ho}$, and in
addition that the system is sufficiently dilute that the
attractive interaction between fermions is likewise less than
$\hbar\omega_{\rm ho}$, it is possible to tune the frequency of
the photoassociation laser so as to only couple fermions in the
shell $n_F$ with Fermi energy ${\cal E}_F$ to molecules in the ground
state of the harmonic trap. We can then ignore all shells other than the
$n_F$-shell for the fermions, and all trap states above the ground state
for the molecules, which are then described in terms of the ground
state bosonic annihilation operator $b$ with single particle
energy ${\cal E}_0$.

Both atomic pairing and photoassociation involve the creation and
annihilation of pairs of atoms, hence it is convenient to
introduce pseudo-spin operators $S_l$~\cite{Kerman61}
for atoms of angular momentum $l$ in the $n_F$-shell as
\begin{eqnarray}
    S_l^+&=&\sum^l_{m=-l}(-1)^{l-m} c_{lm\su}^\dagger
    c_{l-m\sd}^\dagger,\\
    S_l^-&=&\sum^l_{m=-l}(-1)^{l-m} c_{l-m\sd}
    c_{lm\su},\\
    S_l^z&=&\frac{1}{2}\left[\sum_{m,\sigma}c_{lm\sigma}^\dagger
    c_{lm\sigma}-\Omega_l\right]=\frac{1}{2}\left(\hat{n}_l-\Omega_l\right),
    \end{eqnarray}
where $\Omega_l = 2l+1$ and $\hat{n}_l$ is number operator in each
level $l$. (Here and hereafter we have omitted the label $n_F$
from the fermionic operator $c_{n_F;l,m,\sigma}$.) They are easily
seen to obey the SU(2) algebra
    \begin{equation}
    [S_l^+,S_{l'}^-]=2S_l^z \delta_{l,l'}, \,\,\,\,\,\,\,
    [S_l^z,S_{l'}^{\pm}]=\pm S_l^\pm \delta_{l,l'}.
    \end{equation}

Since we need only consider atoms in the Fermi level, the total
number of relevant particles in the system is
    \begin{equation}
    N= n_p + n_b + \nu \equiv M + \nu,
    \label{Ntotal}
    \end{equation}
where
    \begin{equation}
    M= n_p + n_b
    \end{equation}
is the number of molecules ($n_b$) and atomic pairs ($n_p$)
in the $n_F$-shell, or loosely speaking the number of pairs,
and $\nu$ is the number of unpaired atoms in the $n_F$-shell.
In that reduced Hilbert space, a complete set of
states is given by
    \begin{eqnarray}
    &&|n_{l},n_{l^\prime},\cdots,n_{l^{\prime\prime}},n_b;\nu\ket\nonumber\\
    &&\quad =\frac{1}{\sqrt{\mathcal N}}
    (S_l^+)^{n_l}(S_{l^\prime}^+)^{n_{l^\prime}}
    \cdots(S_{l^{\prime\prime}}^+)^{n_{l^{\prime\prime}}}
    (b^\dagger)^{n_b}|\nu\ket,
    \end{eqnarray}
where $\mathcal N$ is a normalization constant.

For a given angular momentum $l$, the possible number of atomic
pairs $n_l$ in the Fermi level is
    $$
    0\leq n_l \leq \Omega_l = 2l+1,
    $$
while the number of molecules is $0\leq n_b \leq N$.

The pseudo-spin operator $S_l^{-}$ annihilates atoms in pairs,
hence any state
$|\nu\ket\equiv|\nu_l,\nu_{l^\prime},\cdots,\nu_{l^{\prime\prime}}\ket$
of unpaired fermions and zero molecules clearly satisfies
    \begin{equation}
    S_l^{-}|\nu\ket=0,\,\,\,\,\,\, \hat{n}_b|\nu\ket=0,
    \label{nu-definition1}
    \end{equation}
with
    \begin{equation}
    \hat{n}_l|\nu\ket=\nu_l|\nu\ket,
    \label{nu-definition2}
    \end{equation}
see Eq. (\ref{Ntotal}). The number operator of bosonic molecules has been
defined by $\hat{n}_b=b^\dagger b$. $\nu_l$ is the number of unpaired fermions of
angular momentum $l$ in each level and is
referred to as {\it seniority}~\cite{Racah} in nuclear physics.

With this formal development at hand, we now turn to the
discussion of the fermionic pairing and of the photoassociation of
atoms into molecules. It is described by the effective Hamiltonian
\begin{equation}
\label{modelH} H=(\hbar \delta+{\cal E}_0) b^\dagger
b+\sum_{l,m,\sigma}{\cal E}_{n_F,l} c_{lm\sigma}^\dagger
c_{lm\sigma} +V_{p} +V_{am},
\end{equation}
where $\delta$ is the two-photon detuning between the Raman lasers
and the internal energy difference between atomic pairs and
molecules, $V_p$ describes atomic pairing and $V_{am}$ accounts
for the photoassociation of atoms into molecules. Before
discussing these two interaction Hamiltonians in detail, we first
evaluate the mean-field lifting of the single-particle energy
degeneracy, ${\cal E}_{n_F} \rightarrow {\cal E}_{n_F,l}$.

In the $s$-wave scattering approximation, valid at $T=0$, atoms of
opposite spin interact via the two-body interaction potential
    \begin{equation}
    V({\bf r}_1 - {\bf r}_2) = \frac{4 \pi \hbar^2 a}{m_f}
    \delta({\bf r}_1 - {\bf r}_2),
    \end{equation}
where $a <0$ is the scattering length, negative for attractive
interactions. In the Thomas-Fermi limit, this results in the atoms
being subjected to the mean-field potential
\begin{equation}
V(r)=\frac{2\pi\hbar^2a}{m_f}\rho(r),
\end{equation}
where the density $\rho(r)$ is given by
\begin{equation}
\rho(r)\simeq \rho_0(1-r^2/R^2_{TF})^{3/2},
\end{equation}
for
    $$
    r\leq R_{\rm TF}=a_{\rm ho}\sqrt{2n_F+3}
    $$
and is zero otherwise. Here $a_{\rm ho}=\sqrt{\hbar/m_f
\omega_{\rm ho}}$ is the oscillator length and
$\rho_0=(2n_F+3)^{3/2}/3\pi^2a_{\rm ho}^3$. The resulting mean-field
energy splitting of the $l$-states within the $n_F$ manifold is
then \cite{Heiselberg}
\begin{eqnarray}
\label{MFshift}
{\cal E}_{n_F,l}&-&{\cal E}_{n_F}\nonumber\\&=&\int\,dr\, r^2V(r)|R_{n_F,l}|^2
\\
&\simeq&\frac{2}{3\pi}\frac{a}{a_{\rm
ho}}(2n_F+3)^{3/2}\hbar\omega_{\rm ho}\left[
\frac{4}{3\pi}-\frac{1}{4\pi}\frac{l(l+1)}{n_F^2}\right]\nonumber,
\end{eqnarray}
where we have used the WKB limit of the radial harmonic oscillator
wave function which is valid for $n_F\gg 1$.
For an atomic system to be dilute, the mean-field shift Eq.(\ref{MFshift})
should be less than the unperturbed energy ${\cal E}_{n}$. This
implies that
\begin{equation}
n_F^{1/2}\frac{|a|}{a_{\rm ho}}\ll 1,
\end{equation}
which is equivalent to the familiar diluteness condition
$\rho_0|a|^3\ll 1$.

It is known that for attractive short range interactions, the potential
$V({\bf r}_1-{\bf r}_2)$ favors the creation of a time reversal
state and lets pairing take place. The Hamiltonian $V_p$ describes
this pairing correlation by including only the correlations for time reversal
pair states
    \begin{equation}
    |L=0,M=0;ll\ket=\sum_{m=-l}^l (l m ,l -m|0 0)|l,m\ket|l,-m\ket,
    \end{equation}
where
    \begin{equation}
    (l m,l -m|0 0)=(-1)^{l-m}(2l+1)^{-1/2}
    \label{CB-coeff}
    \end{equation}
is a Clebsch-Gordan coefficient. Assuming for simplicity that the
radial part $g$ of the pairing interaction is independent of the
angular momenta $l$ and $l'$ of the atomic pairs involved in the
interaction, $V_p$ reads explicitly
    \begin{equation}
    \label{pinteraction}
    V_{p}=-g\sum_{l,m}\sum_{l^\prime,m^\prime}
    (-1)^{l-m+l^\prime-m^\prime}c_{lm\su}^\dagger c_{l-m\sd}^\dagger
    c_{l^\prime m^\prime\sd}c_{l^\prime
    -m^\prime\su},
    \end{equation}
where the terms $(-1)^{l-m}$ and $(-1)^{l^\prime-m^\prime}$ come
from the Clebsch-Gordan coefficients (\ref{CB-coeff}) and hence
account for the angular part of the wave functions.
Strictly speaking, the coupling constant $g_{l,l'}$ should be
determined by the spatial integral
\begin{equation}
g_{l,l^\prime}=\frac{4\pi\hbar^2 |a|}{m_f}\int dr\,r^2\,
R_{n_F,l}^2(r)R_{n_F,l^\prime}^2(r). \label{g-spatial}
\end{equation}
However for our present purpose it is sufficient to estimate $g$ by
replacing the spatial integral in Eq. (\ref{g-spatial}) by $\bra
r_{n_F}^2 \ket^{-3/2}$, where for a pure harmonic oscillator state
in the $n_F$-shell the mean square radius $\bra r_{n_F}^2 \ket$ is
given by $(2n_F+3)a_{\rm ho}^2$ from the virial theorem. This
gives
\begin{equation}
g\sim \frac{4\pi\hbar^2 |a|}{m_f} (2n_F+3)^{-3/2}a_{\rm ho}^{-3}
\sim \frac{|a|}{a_{\rm ho}}n_F^{-3/2}\hbar\omega_{\rm ho}.
\end{equation}

In this model, the strongest pairing occurs for degenerate
energies, ${\cal E}_{l}={\cal E}_{l^\prime}$ for all $l,l^\prime$.
We will see that all fermions in the $n_F$-shell are then
coherently paired, the pairing energy being proportional to the
degeneracy factor $\Omega_{n_F}$. In the dilute gas limit, which
is equivalent to $g\Omega_{n_F}\sim g n_F^2 \ll \hbar\omega_{\rm
ho}$, the pairing takes then place on a single shell, see
Ref.~\cite{Heiselberg} for more details.

Turning finally to the photoassociation of atomic pairs into
molecules and the reverse process of photodissociation of
molecules back into atoms, we note that it is possible to neglect
all processes involving atoms other than those in the $n_F$-shell
provided that their characteristic frequency, the product $\chi
\sqrt{ \langle n_b \rangle}$ of the photoassociation coupling
constant, $\chi$, and the square root of the mean number of
molecules, remains small compared to the frequency separation
between neighboring shells of the trap. This condition is well
fulfilled for typical laser strengths and trap depths, in which
case $\chi \sqrt{ \langle n_b \rangle} \simeq 10^2 n_F^{-1/2}
\,{\rm s}^{-1} \ll \omega_{\rm ho} \simeq 10^3 \,{\rm s}^{-1}$.
In this case, the atom-molecule coupling can be approximated by the Hamiltonian
\begin{equation}
\label{aminteraction}
V_{am}=\sum_{l,m}(-1)^{l-m}\left[\chi
c_{lm\su}^\dagger c_{l-m\sd}^\dagger b +\chi^* b^\dagger
c_{l-m\sd} c_{lm\su}\right],
\end{equation}
where we only include the coupling between time-reversal atomic
pair and bosonic molecules in the ground state of the harmonic
trap. The photoassociation coupling constant $\chi$ is
proportional to the far off-resonant two-photon Rabi frequency
associated with two nearly co-propagating lasers with frequencies
$\omega_1$ and $\omega_2$~\cite{Wynar00},
\begin{eqnarray}
\label{coupling} \chi&=&\frac{\chi_0}{\sqrt{4\pi}} \int\, dr\,
r^2\, R^{(b)}_{0,0}(r) R_{n_F,l}^2(r)
\nonumber\\
&\sim& \frac{\chi_0}{n_F^{3/2} a_{\rm ho}^{3/2}},
\end{eqnarray}
where $R^{(b)}_{0,0}$ is a ground-state wave function of molecules
of mass of $2m_f$ with a spatial width about $a_{ho}$,
and in the second equality we
have again replaced the radial wave function of the atoms by 
$\bra r^2_{n_F} \ket^{-3/2}$.
From the coupling constant between $^{87}$Rb$_2$ molecules and $^{87}$Rb
atoms of Ref.~\cite{superchemistry} we estimate the
photoassociation coupling constant to be of the order of
$\chi_0=7.6 \times 10^{-7}$ m$^{3/2}$s$^{-1}$.

In terms of the pseudo-spin operators, and with Eqs.
(\ref{pinteraction}) and (\ref{aminteraction}), the total model
Hamiltonian Eq.~(\ref{modelH}) finally reads
\begin{eqnarray}
\label{spinH} H&=&(\hbar \delta+{\cal E}_0) b^\dagger b+\sum_l
2{\cal E}_{n_F,l}(S_l^z+\Omega_l/2)\nonumber\\
&+&\chi\sum_l (S_l^+ b+b^\dagger
S_l^-)-g\sum_{l,l^\prime}S_l^+S_{l^\prime}^-.
\end{eqnarray}
This Hamiltonian clearly conserves the spin operators ${\bf
S}_l^2$. Applying this operator on the state $|\nu\rangle$ we have
\begin{eqnarray*}
{\bf S}_l^2|\nu\ket&=&S_l(S_l+1)|\nu\ket \\
&=& \{S_l^+S_l^-+S_l^z(S_l^z-1)\}|\nu\ket \\
&=& (\Omega_l/2-\nu_l/2)(\Omega_l/2-\nu_l/2+1)|\nu\ket,
\end{eqnarray*}
where we have used the identity ${\bf
S}_l^2=S_l^{+}S_l^{-}+S_l^{z}(S_l^{z}-1)$ and Eqs.
(\ref{nu-definition1}) and (\ref{nu-definition2}), so that
    $$
    S_l=\frac{1}{2}(\Omega_l-\nu_l).
    $$
This allows us to identify the operator
    \begin{equation}
    \hat{M}=\sum_l \left(S^z_l+S_l\right)+\hat{n}_b,
    \end{equation}
as the operator giving the total number of fermionic pairs and
molecules, or loosely speaking the ``pair number'' operator, which
is easily seen to also be a conserved quantity.

In the limit $g \to 0$ the Hamiltonian~(\ref{spinH}) reduces to
the Tavis-Cummings Model~\cite{Tavis,Bogoliubov} of quantum
optics, while for $\chi \to 0$, it becomes the Pairing Model
~\cite{Richardson,Dukelsky}, for which Richardson first gave an
exact solution in the context of nuclear physics. We also mention
a recent family of exactly solvable models of atom-molecule
proposed in Ref.~\cite{AMHamiltonian}.

The discussion of the following sections concentrates specifically
in that situation where the mean-field energy shift in
single-particle energies is smaller than the photoassociation
coupling. In this degenerate model, the single-particle energies
of all atoms in the $n_F$-shell can then be taken to be equal,
${\cal E}_{n_F,l}= {\cal E}_F$, and the Hamiltonian (\ref{spinH})
simplifies to
    \begin{equation}
    \label{DGmodel}
    H=-\hbar \omega\left(S^z+S\right)
    +\hbar\chi\left(S^{+}b+b^\dagger S^{-}\right)-\hbar gS^{+}S^{-},
    \end{equation}
where we have introduced the total spin operator
    \begin{equation}
    {\bf S}=\sum_l {\bf S}_l,
    \end{equation}
$\omega=\delta+({\cal E}_0-2{\cal E}_F)/\hbar$, $\Omega_{n_F}=\sum_l \Omega_l$, $\nu=\sum_l \nu_l$, so that the total spin is
    \begin{equation}
    S= \sum_l S_l=\frac{1}{2}\left (\Omega_{n_F} - \nu \right ).
    \label{total spin}
    \end{equation}
In Eq.~(\ref{DGmodel}), we have neglected constant terms
proportional to the conserved quantity $M$.

Assuming an oscillator length $a_{\rm ho} =3.2\times 10^{-6}$ m
for the mass of $^6$Li, $a=-114$ nm for its scattering length, and
$\omega_{\rm ho}=1000$ s$^{-1}$, then $\chi n_F^{3/2} \simeq 10^{2}$
s$^{-1}$.  The validity of the degenerate model, $|{\cal
E}_{n_F,n_F}- {\cal E}_{n_F,l=0}| < \chi n_F$ requires $n_F
\lesssim 10$, which corresponds to a total number of fermions 
$N_{n_F} \lesssim 10^3$.

\section{Ground State}

In this section, we discuss the dependence of the ground state of
the model Hamiltonian (\ref{DGmodel}) on the ratio of number of
paired fermions and molecules, $M=n_p+n_b,$ to the degeneracy
$\Omega$ of the Fermi level (we drop the subscript `$n_F$' for
notational clarity from now on). We identify qualitatively
different types of ground states, a pair-dominated ground state
and a molecular-dominated one, as a function of the parameter
    \begin{equation}
    \kappa \equiv \frac{\omega}{\chi \sqrt{\Omega}},
    \end{equation}
where $\omega = \delta +({\cal E}_0-2 {\cal E}_F)/\hbar$ is the
photoassociation frequency detuning.

\subsection{Pairing Model}

For a total number of particles $N=2M+\nu \leq \Omega$ , which
corresponds to the Fermi energy shell less than half-filled since
there are two hyperfine atomic states involved, the seniority
$\nu$ can take the values
\begin{equation}
  \nu = \cases{0,2,4,\dots,N & ({\em N} \, even) \cr
             1,3,5,\dots,N & ({\em N} \, odd), \cr}
\end{equation}
while for $\Omega < N \leq 2\Omega$, corresponding to a shell more
than half filled, the permissible values of $\nu$ are
\begin{equation}
  \nu = \cases{0,2,4,\dots,2\Omega-N & ({\em N} \, even) \cr
             1,3,5,\dots,2\Omega-N & ({\em N} \, odd). \cr}
\end{equation}

The degenerate Pairing Model of Refs.~\cite{Schuck,Walecka} is
obtained by neglecting the photoassociation coupling, $\chi=0$, in
the Hamiltonian~(\ref{DGmodel}). In this case, the total energy is
given as a function of seniority $\nu$ and the total number $N$ of
fermions by
\begin{equation}
E_{\rm pm}(N,\nu)=-\omega N -\frac{g}{4}(N-\nu)(2\Omega-\nu-N+2),
\end{equation}
and it is minimized for $\nu=0$ or $1$.

For an even particle number, the ground state corresponds
therefore to all pseudo-spins aligned, $S = \Omega/2$,
    \begin{equation}
    E_{\rm pm}(N,\nu=0)=-\omega N -gN(2\Omega -N + 2).
    \end{equation}
The first excited state corresponds to $\nu=2$ unpaired atoms, or
$S=\Omega/2-1$, and its energy is
\begin{equation}
\label{gapenergy} E_{\rm pm}(N,\nu=2)-E_{\rm pm}(N,\nu=0)=g\Omega.
\end{equation}
Thus, the energy needed to break up an atomic pair into two
un-paired fermions is independent of the number of fermions in the
$n_F$-shell. This energy is consistent to Bogoliubov
quasi-particle energy based on BCS variational state except for
corrections of relative order $1/\Omega$.

\subsection{Photoassociation}

In the presence of photoassociation, $\chi \neq 0$, the
eigenstates of the system consist of a coherent superposition of
atoms and molecules. For fixed $N$, we can classify them by the
total spin $S=(\Omega-\nu)/2$, each manifold consisting of $(M+1)$
eigenstates, where $M=(N-\nu)/2 = n_p + n_b$ is the total number
of pairs.

For positive detuning, $\omega>0$, the ground state is a pure
fermionic state, i.e., $N=2n_p+\nu$. On the other hand, for a
large negative detuning energy, $\omega<0$, and even particle
number, the ground state is reached when all particles are
molecules, corresponding to a zero seniority, or maximum spin
state.

\begin{figure}
%  \begin{minipage}{0.45\textwidth}
    \begin{center}
      \includegraphics[width=7.5cm,clip]{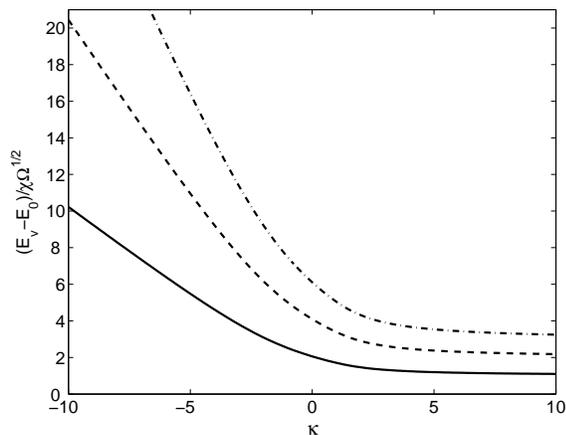}
    \end{center}
    \caption{Excitation energies, in units of $\chi \sqrt{\Omega}$,
     relative to the ground-state energy
     for the seniority states
     $\nu=2$ (solid line), $\nu=4$ (dashed line), and $\nu=6$
     (dash-dotted line), as a function of the dimensionless
     parameter $\kappa=\omega/\chi\sqrt{\Omega}$
     for $\eta=g\sqrt{\Omega}/\chi=1$ and $N=\Omega=120$.}
    \label{gsfig1}
\end{figure}

Figure~\ref{gsfig1} shows the energies $E_\nu$ of a few seniority
states relative to that of the state $\nu=0$ as a function of
$\kappa = \omega/\chi\sqrt{\Omega}$. This example is for $\eta=
g\sqrt{\Omega}/\chi=1$ and a half-filled Fermi energy shell,
$N=\Omega=120$, corresponding to $n_F=14$, and results from a
direct numerical diagonalization of the Hamiltonian. It shows that
over the wide range of $\kappa$ considered here, the ground state
is always the maximum spin manifold $\nu=0$. For large negative
detunings, the energy differences $E_\nu-E_0$ are $\nu/2$ times
the single-particle energy difference between a molecule and two
fermions, and increase with $|\omega|$. For large positive
detunings, on the other hand, the energies of the excited states
approach the values $\nu/2$ times $g\Omega$, see
Eq.~(\ref{gapenergy}). In the crossover region where the nature of
the ground state changes from atomic to molecular, the ground
state is a coherent superposition of atoms and molecules that is
likewise the maximum spin state. We have verified that in the
absence of pairing interaction, $g=0$, the ground state is
likewise the state of maximum spin (not shown in figure).

In the following we concentrate on the state of maximum total
spin, $|S=\Omega/2\ket$, for an even number of fermions, $N=2M$.
The eigenstates of the atom-molecule system in the spin manifold
$S=\Omega/2$ have the general form
\begin{eqnarray}
|\Phi^\lambda_S\ket&=&\sum_{n_b=0}^{M} C_{\lambda}(n_p)
|S=\Omega/2,S^z=-S+n_p\ket_f \nonumber\\&&\otimes |n_b=M-n_p\ket_b,
\end{eqnarray}
where $n_p$ denotes a number of fermionic pairs and
$\lambda=0,1,\cdots M$ represent eigenmodes of the system with
eigenenergies $E_\lambda$. We have found these eigenstates and the
associated eigenenergies by direct diagonalization of the
Hamiltonian~(\ref{DGmodel}).

\begin{figure}
  \begin{minipage}{0.45\textwidth}
    \begin{center}
      \includegraphics[width=7.5cm,clip]{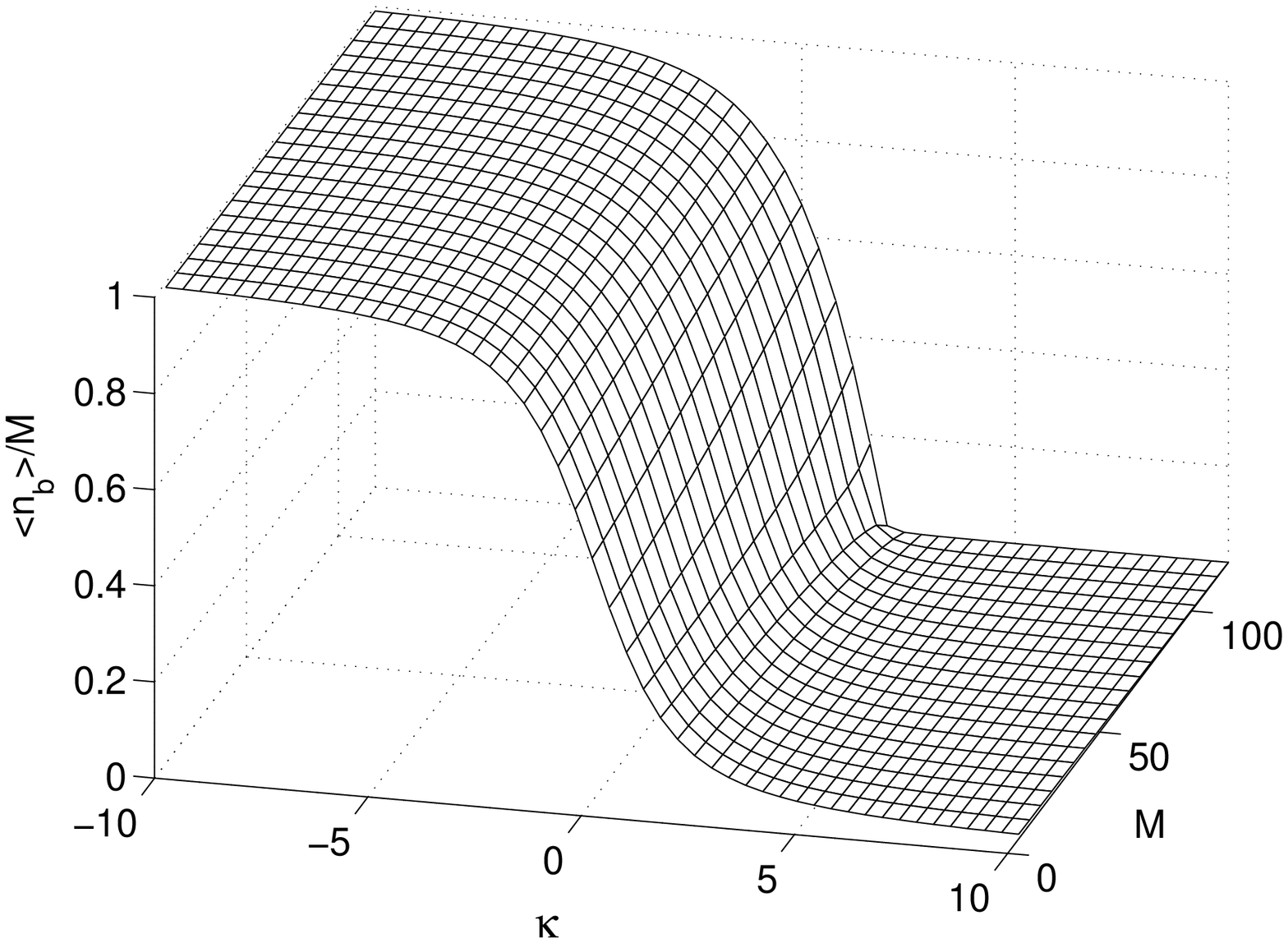}
    \end{center}
    \caption{Normalized average number of molecules in the ground state
    as a function of $\kappa$ and $M$ for $\Omega=120$ and $g=0$.}
    \label{gsfig2}
  \end{minipage}
\hspace{4mm}
  \begin{minipage}{0.45\textwidth}
    \begin{center}
      \includegraphics[width=7.5cm,clip]{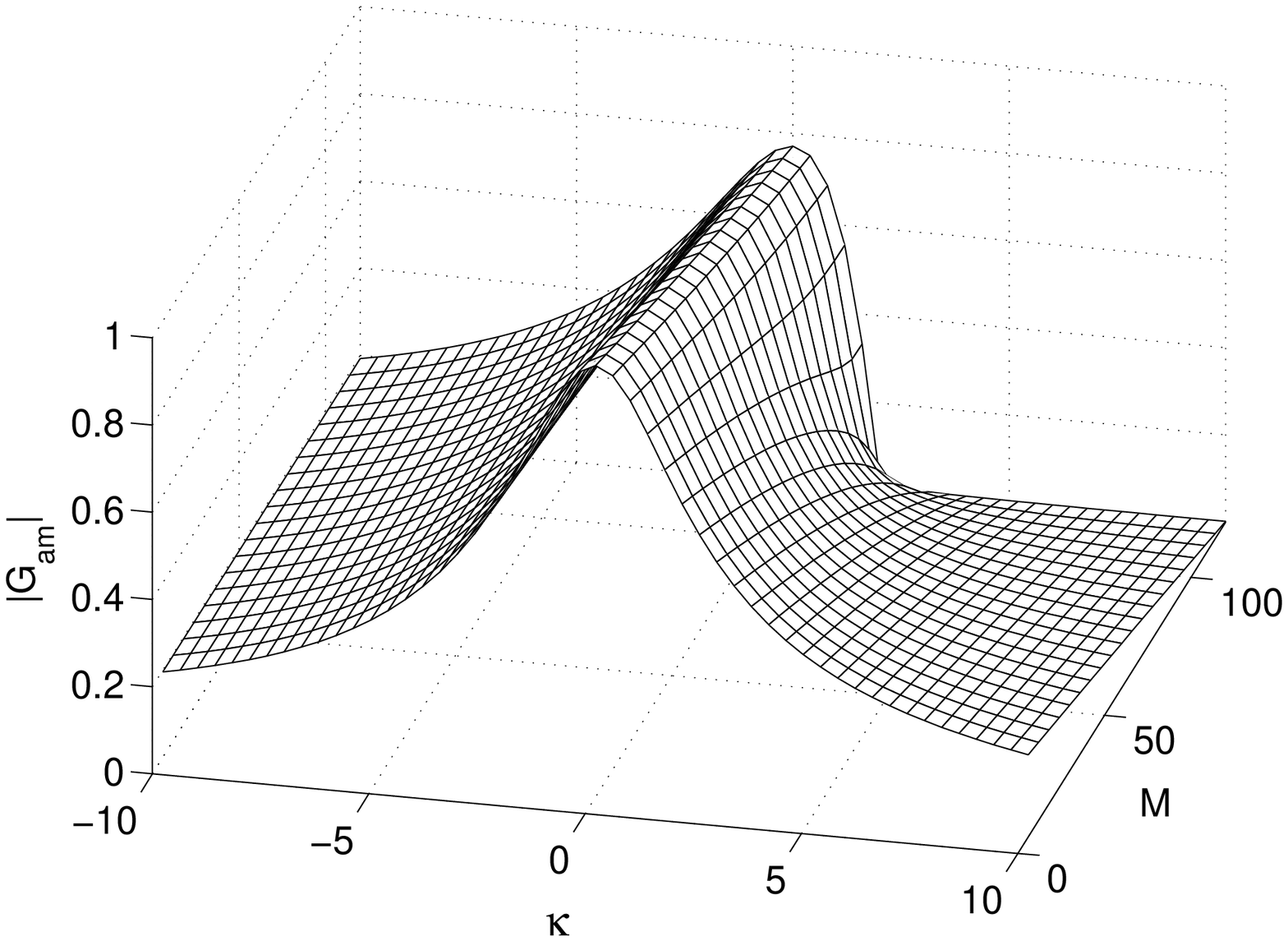}
    \end{center}
    \caption{Normalized joint coherence function, $|G_{am}|$,
      as functions of $\kappa$ and $M$ for $\Omega=120$ and for $g=0$.}
    \label{gsfig3}
  \end{minipage}
\end{figure}

The average number of molecules in the ground state is shown in
Fig.~\ref{gsfig2} as a function of
$\kappa=\omega/\chi\sqrt{\Omega}$ and $M$ for $g=0$ and
$\Omega=120$. As expected, for large positive detuning $\omega>0$,
the ground state population consists almost exclusively of atoms,
while it is mostly made up of molecules for large negative
detunings. In the crossover region around $\omega=0$, the ground
state consists of a coherent superposition state between molecules
and atomic pairs.

It is possible to characterize this superposition in terms of the
normalized joint coherence function
    \begin{equation}
    G_{am} = \frac{2 \langle S^+ b\rangle}{M\sqrt{\Omega}},
    \end{equation}
which is shown in Fig.~\ref{gsfig3} as a function of $\kappa$ and
the number of pairs $M$. The joint coherence of the atomic and
molecular fields shows a remarkable enhancement in the crossover
region as well as a change in shape as a function of $M$, due to
the nonlinearity of the atom-molecule
coupling~(\ref{aminteraction}). This dependence on the filling
factor $M/\Omega$ can be understood more quantitatively by
considering the two limiting cases $M/\Omega\ll 1$ and $M/\Omega
\simeq 1$.

\subsubsection{$M/\Omega \ll 1$ --- mapping on a linear coupled-boson
system}

In the limit of small filling factors, it is convenient to
describe the system in terms of the Holstein-Primakoff mapping
~\cite{HP40} of the $SU(2)$ generators
$S^+=(S^{-})^\dagger$ and $S^{z}$ in terms of bosonic operators.
According to this mapping, the Hilbert space of the group $SU(2)$
is carried by a subspace of the bosonic Fock space given by the
bosonic vacuum $|0\rangle_d$ and the bosonic operators $d$ and
$d^\dagger$, with
    \begin{equation}
    [d,d^\dagger]=1, \qquad d|0\ket_d=0,
    \end{equation}
The bosonic space being spanned by the $n_d-$boson states
    \[
    |n_d\ket_d=\frac{1}{\sqrt{n_d !}}(d^\dagger)^{n_d}|0\ket_d \qquad
    {\rm for}\,\, n_d=0,1,2,\dots,M.
    \]
Since we restrict our considerations to the subspace characterized
by the angular momentum quantum number $S=\Omega/2$, we can map
the operators $S^{\pm}$ and $S^{z}$ as
\begin{eqnarray}
    \label{HPmapping}
    S^{+} &\to& \sqrt{\Omega} d^\dagger
    \sqrt{1-\frac{d^\dagger d}{\Omega}},\nonumber \\
    S^{-} &\to& \sqrt{\Omega} \sqrt{1-\frac{d^\dagger d}{\Omega}} d,\nonumber \\
    S^z &\to& -\frac{\Omega}{2}+d^\dagger d.
    \end{eqnarray}

In the limit $M/\Omega\ll 1$, we only need a lowest-order of the
operators (\ref{HPmapping}), that is, replace the square roots by
1. The Hamiltonian~(\ref{DGmodel}) reduces then to that of a
linearly coupled two-mode boson system,
\begin{equation}
    \label{linearboson}
    H\to H_{\rm linear}=-(\omega+g\Omega)d^\dagger
    d +\chi\sqrt{\Omega}(d^\dagger b +b^\dagger d),
    \end{equation}
with the constraint $M=n_d+n_b$.

This Hamiltonian can be diagonalized by the transformation
    \begin{eqnarray}
    c_{-}^\dagger&=&\cos{\theta}d^\dagger-\sin{\theta}b^\dagger\nonumber \\
    c_{+}^\dagger&=&\sin{\theta}d^\dagger+\cos{\theta}b^\dagger \nonumber,
    \end{eqnarray}
with $\cot{2\theta}=(\omega+g\Omega)/2\chi\sqrt{\Omega}$,
,$0\le\theta\le 2\pi$, to give
\begin{equation}
\label{lineardiagonal} H_{\rm linear}=\epsilon_{-}c_-^\dagger
c_-+\epsilon_{+}c_+^\dagger c_+,
\end{equation}
with energies
    \begin{eqnarray}
    \epsilon_-&=&-(\omega+g\Omega)\cos^2{\theta}
    -2\chi\sqrt{\Omega}\cos{\theta}\sin{\theta}, \nonumber \\
    \epsilon_+&=&-(\omega+g\Omega)\sin^2{\theta}
    +2\chi\sqrt{\Omega}\cos{\theta}\sin{\theta}. \nonumber
    \end{eqnarray}
Since $\epsilon_-<\epsilon_+$ for $\chi > 0$, the ground state is
$(1/\sqrt{M !})(c_-^\dagger)^M|0\ket_d|0\ket_b$.

\subsubsection {$M/\Omega \simeq 1$ --- Mapping on a binary
atomic-molecular condensate}

Tikhonenkov and Vardi~\cite{Vardi} showed for in the homogeneous
case that when the total number of pairs is equal to the available
momentum states of the fermions, the system of fermionic atoms and
bosonic molecules can be mapped onto a two-mode atomic-molecular
BEC system~\cite{AMBEC1,AMBEC2,Zhou}. The corresponding situation
in our case occurs for $M=\Omega$, with an additional
atom-molecule two-body collision term required in addition.

To show how this works we introduce the two-mode Hamiltonian of a
two-component condensate of atoms and molecules,
\begin{equation}
\label{amhamiltonian}
H_{am}=-\omega b_m^\dagger b_m + \frac{\chi}{2}(b_m^\dagger b_a b_a+h.c.)
-\frac{g}{2}b^\dagger_m b_m b^\dagger_a b_a,
\end{equation}
where $b_a^\dagger$ and $b_m^\dagger$ are the bosonic creation
operators for the atomic and molecular modes, respectively.
Clearly the total number of particles $N_{am}=n_a+2n_m$ is
conserved, where $n_a$ and $n_m$ are the number of atoms and
molecules, respectively. For $N_{am}$ even, a general state of the
system can be expressed as
\begin{eqnarray}
|\Phi_{am}\ket&=&\sum_{m_a=0}^{M}C(m_a)|n_a=2m_a\ket\otimes
|n_m=M-m_a\ket \nonumber \\
&=& \sum_{m_a=0}^{M}C(m_a)|2m_a,n_m \ket,
\end{eqnarray}
with $M=N_{am}/2$. In this representation, the matrix form of the
Hamiltonian~(\ref{amhamiltonian}) is
     \begin{eqnarray*}
       \label{amMH}
       &&\bra 2m_a;n_m |H_{am}| 2m_a;n_m\ket\\
       &&\qquad =-\omega(M-m_a)-gm_a(M-m_a)\\
       &&\bra 2(m_a-1);n_m+1 |H_{am}| 2m_a;n_m\ket\\
       &&\qquad=\bra2m_a;n_m|H_{am}|2(m_a-1);n_m+1\ket \\
       &&\qquad=\chi \sqrt{m_a(m_a-1/2)(M-m_a)}.
     \end{eqnarray*}
Similarly, for our model the matrix form of the
Hamiltonian~(\ref{DGmodel}) in the ``pair number'' representation
is
    \begin{eqnarray*}
      \label{fbMH}
      &&\bra n_b;n_p|H|n_b;n_p\ket\\
      &&\qquad =-\omega(M-n_b)-g (n_b+\Delta M+1)(M-n_p)\\
      &&\bra n_b-1;n_p+1 |H| n_p;n_b\ket\\
      &&\qquad =\bra n_p;n_b |H|n_b-1;n_p+1\ket\\
      &&\qquad =\chi \sqrt{n_b(n_b+\Delta M+1)(M-n_b)}.
    \end{eqnarray*}
where
    \begin{equation}
    \Delta M=\Omega-M ,
    \end{equation}
with $\Delta M>0$. In the limit $M\simeq \Omega$, that is, when we
can neglect $\Delta M/M \ll 1$ and other terms of order of
$1/\Omega$, these two Hamiltonians are same under the
transformations $m_a\leftrightarrow n_b$ and $n_m \leftrightarrow
n_p$.

\subsubsection{Intermediate regime}

\begin{figure}
    \begin{center}
      \includegraphics[width=7.5cm,clip]{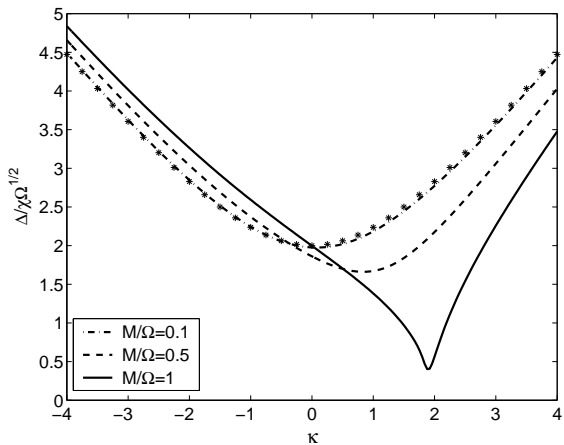}
    \end{center}
    \caption{Lowest excitation energies as a function of
      $\kappa$ for $M/\Omega=1$ (solid line)
      and $M/\Omega=0.5$ (dashed line), and $M/\Omega=0.1$ (dash-dotted line),
      at fixed pair number $M=120$, where $g=0$. The asterisks correspond
      to a linear approximation for small $M/\Omega$ (see text).}
    \label{gsfig4}
\end{figure}

The behavior of the system apart from these two limiting cases
deviates from both models. The most striking difference between
these regimes appears in the energy, $\Delta$, of the first
excited state. Figure~\ref{gsfig4} shows
$\Delta/\chi\sqrt{\Omega}$ in the absence of pairing interaction
for $M/\Omega=1$ (solid line), $M/\Omega=0.5$ (dashed line), and
$M/\Omega=0.1$ (dash-dotted line) as a function of $\kappa =
\omega/\chi\sqrt{\Omega}$ for fixed $M=120$. In the case of
$M/\Omega=0.1$, $\Delta$ agrees well with the one-particle energy
difference $\epsilon_+-\epsilon_-=\sqrt{\omega^2+ 4\chi^2\Omega}$
of the linear coupled-boson model~(\ref{lineardiagonal}) (shown as
asterisks in the figure). Increasing $M/\Omega$ shifts the
location of the minimum of the energy gap and reduces the value of
its minimum. For $M=\Omega$, finally, the minimum gap approaches
zero at $\kappa\simeq 2$, consistently with a transition point in
the atom-molecule condensate system~\cite{AMBEC1,Zhou}. This
result is indicative of the appearance of a quantum phase
transition in the limit $M \to \infty$.

\subsection{The role of the pairing interaction}

\begin{figure}
  \begin{minipage}{0.45\textwidth}
    \begin{center}
      \includegraphics[width=7.5cm,clip]{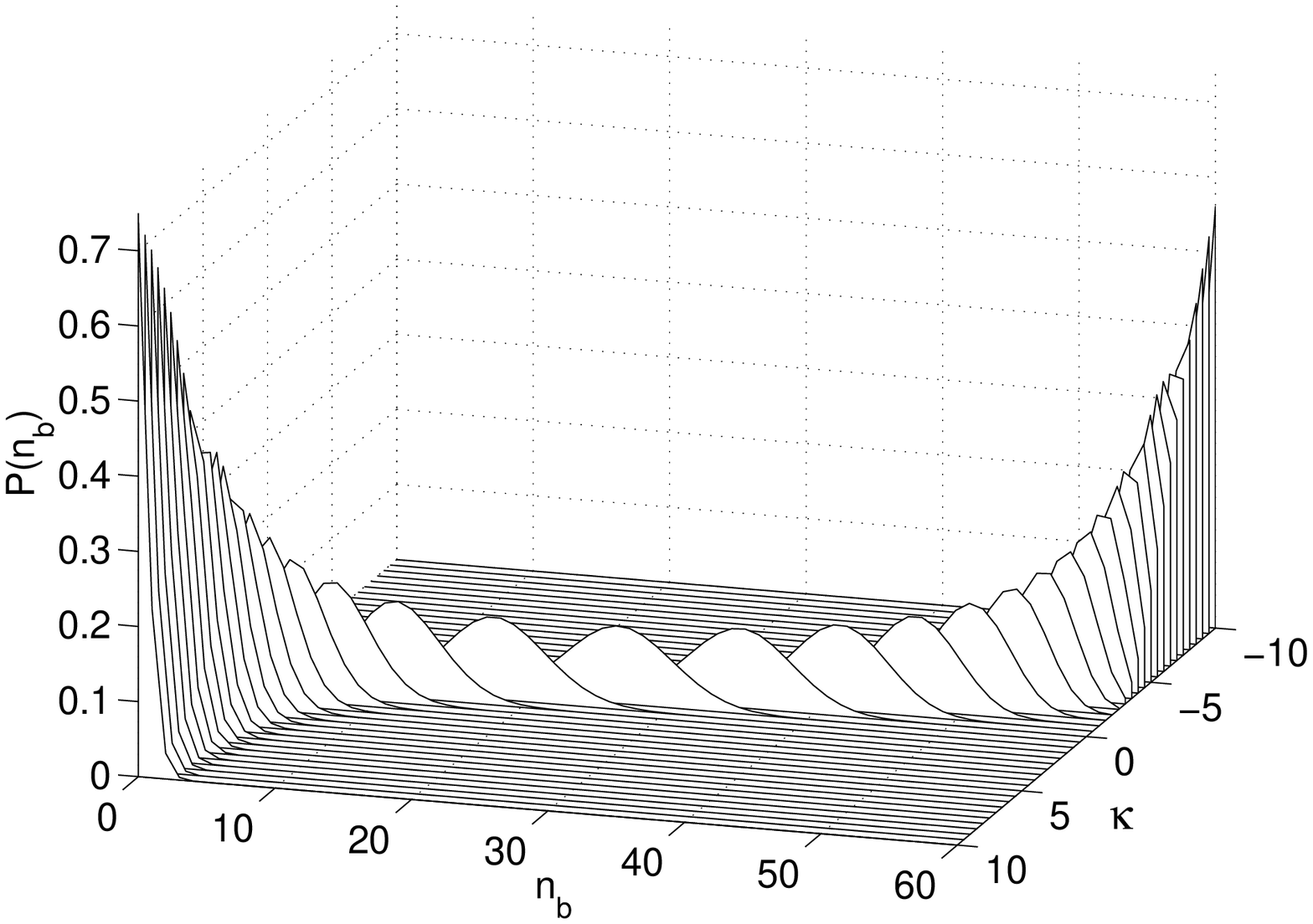}
    \end{center}
    \caption{Molecule statistics $P(n_b)$ as a function of
      the dimensionless parameter $\kappa$ in the absence of
      pairing interaction, $g=0$, for $2M=\Omega=120$.}
    \label{gsfig5}
  \end{minipage}
\hspace{4mm}
  \begin{minipage}{0.45\textwidth}
    \begin{center}
      \includegraphics[width=7.5cm,clip]{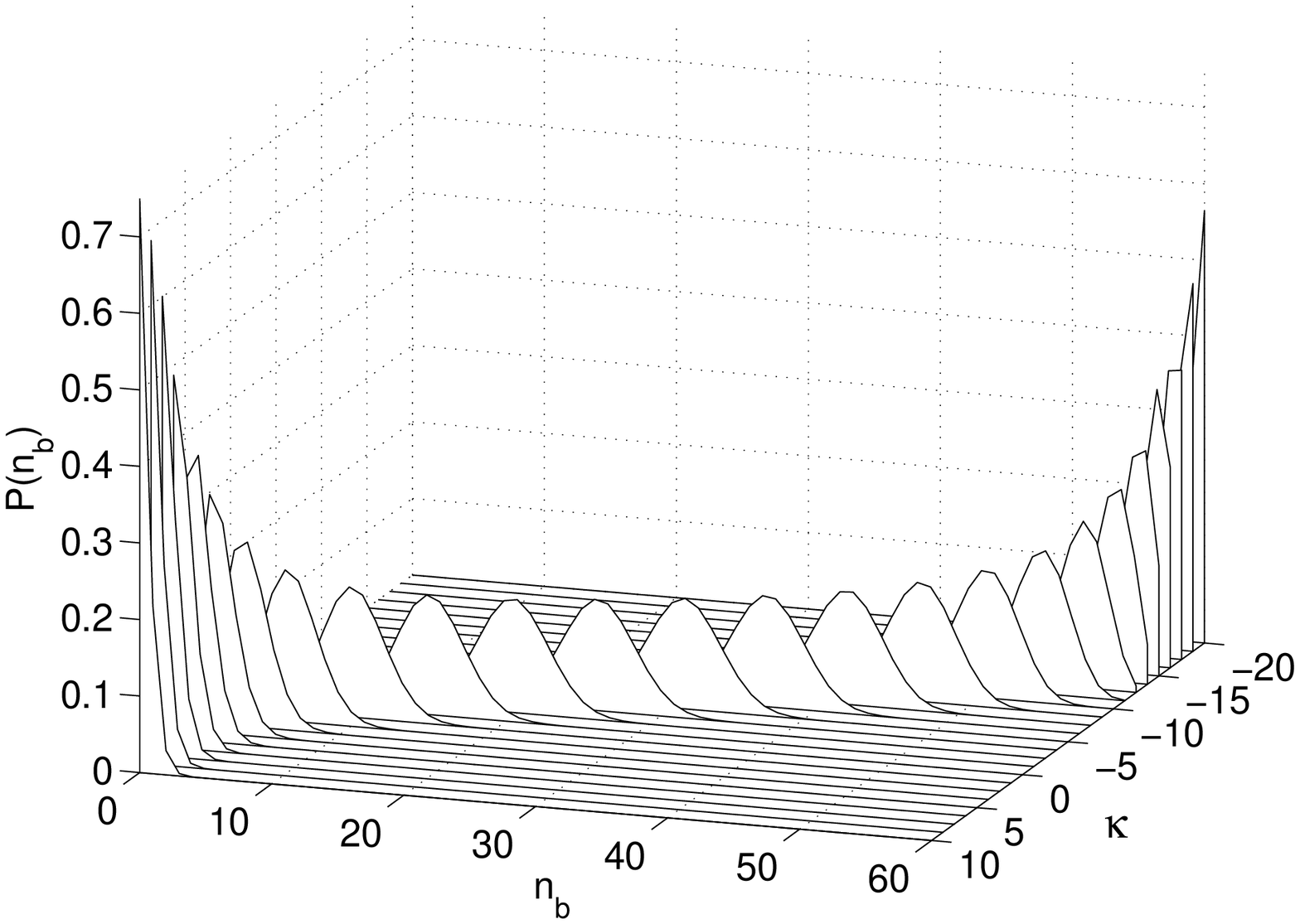}
    \end{center}
    \caption{Molecule statistics $P(n_b)$ as a function of
    the dimensionless parameter $\kappa$ in the presence of
      pairing interaction, $\eta=g\sqrt{\Omega}/\chi=10$, for $2M=\Omega=120$.}
    \label{gsfig6}
  \end{minipage}
\end{figure}

We now examine in more detail the ground-state statistics of the
molecular field in the presence of pairing interaction $V_{p}$.
Since the cases of $M\ll \Omega$ and $M\simeq \Omega$ can be
mapped onto relatively well-known systems, we present results for
the situation of a half-filled shell, $M/\Omega=0.5$, only. Figures
\ref{gsfig5} and \ref{gsfig6} show the probability $P(n_b)$ of
having $n_b$ molecules in the trap, the molecule statistics, for
$2M=\Omega=120$ as a function of the dimensionless parameter
$\kappa$, with $g=0$ and $\eta=g\sqrt{\Omega}/\chi=10$,
respectively. There are several quantitative differences between
the two cases: the transition from an atomic to a molecular ground
state is shifted by the pairing interaction, and the width of the
crossover region in detuning space is significantly broader.

This behavior can be understood by noting that the pairing
interaction gives rise to an additional detuning effect depending
on the number of molecules, as shown by the diagonal terms of the
Hamiltonian~(\ref{fbMH}) with $M=\Delta M=\Omega/2$. This
interaction leads to number-dependent energy shifts and dephasing
between different molecular number states. A similar effect has
been studied in the context of a Jaynes-Cummings-like description
of photoassociation and has been referred as ``nonlinear detuning''
~\cite{search}.

We can estimate the position of the ``resonance'' point
$\omega_{\rm res}(g)$ where the ground state goes from being
molecular to atomic in nature by taking $\bra
n_b=M/2+1;n_p=M/2-1|H|n_b=M/2+1;n_p=M/2-1\ket =\bra
n_b=M/2;n_p=M/2|H|n_b=M/2;n_p=M/2\ket$. This gives
\[
\omega_{\rm res}(g)=-g\frac{\Omega}{2},
\]
where we have neglected a term of order $1/\Omega$.

\begin{figure}
  \begin{minipage}{0.45\textwidth}
    \begin{center}
      \includegraphics[width=7.5cm,clip]{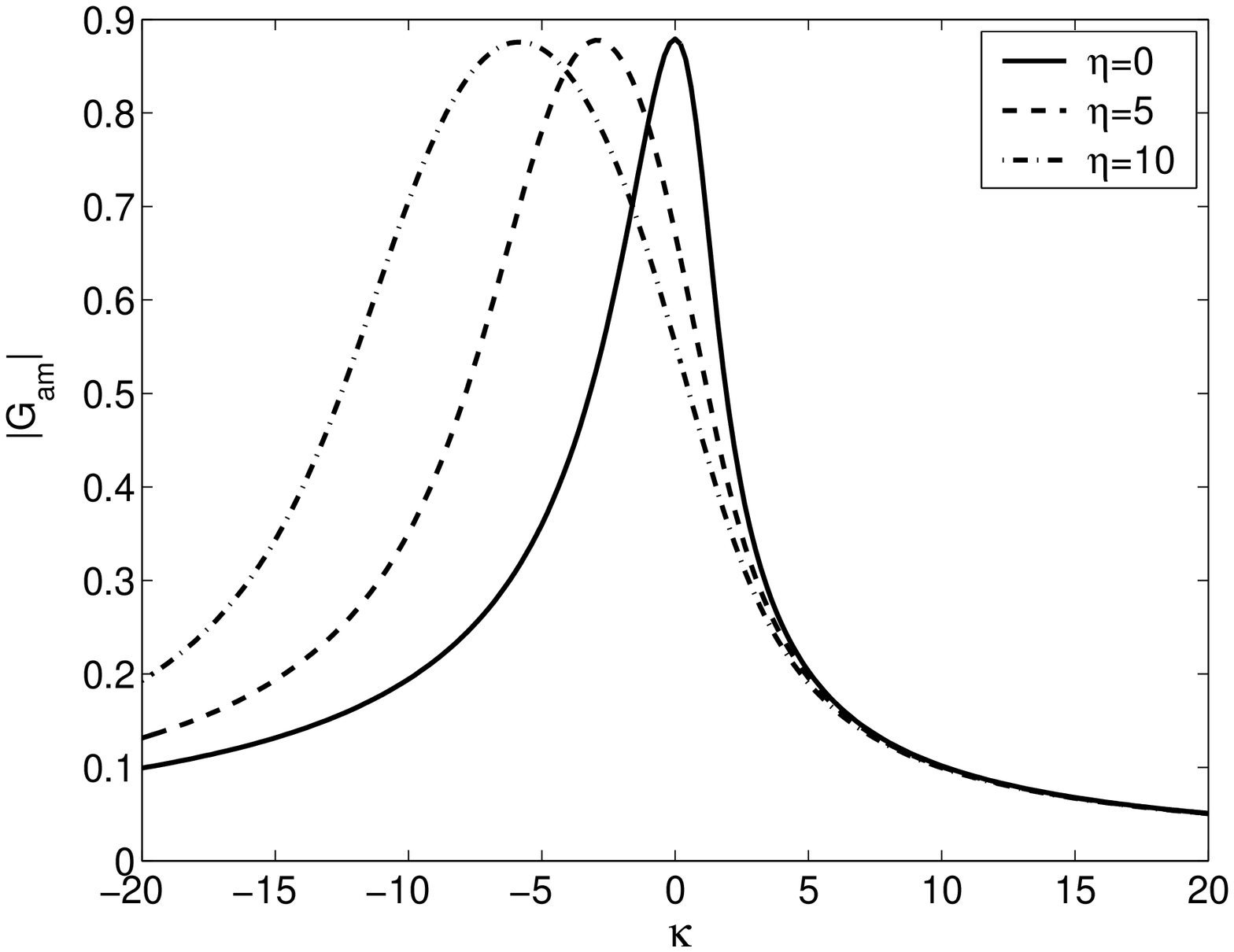}
    \end{center}
    \caption{Normalized joint coherence, $|G_{am}|$, of the ground state
    as a function of $\kappa$
    for three values of the pairing interaction strengths
    $\eta=g\sqrt{\Omega}/\chi$,
    indicated in the insert.}
    \label{gsfig7}
  \end{minipage}
\hspace{4mm}
  \begin{minipage}{0.45\textwidth}
    \begin{center}
      \includegraphics[width=7.5cm,clip]{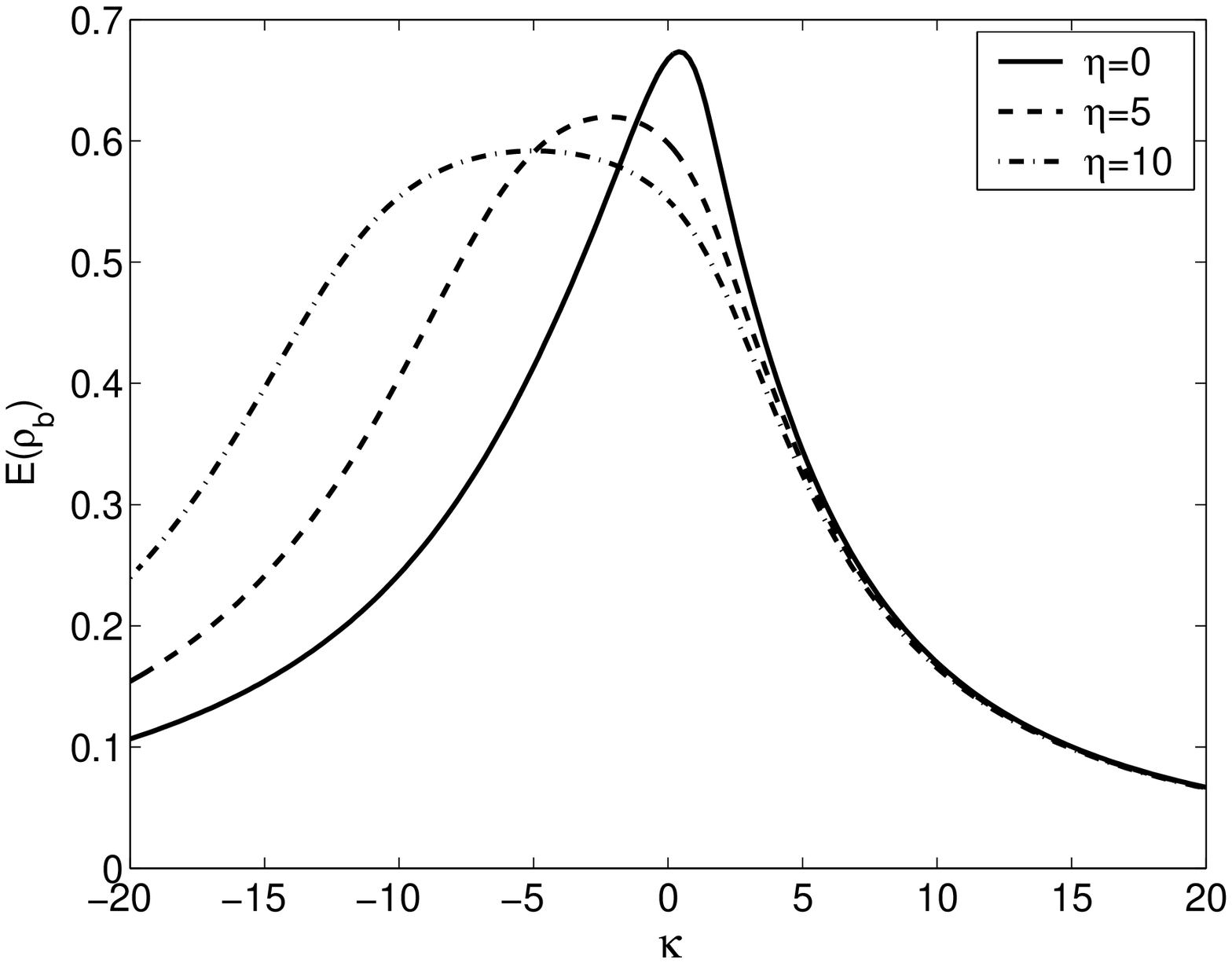}
    \end{center}
    \caption{Entanglement entropy, $E(\rho_b)$, of the ground
    state, in units of the maximum entanglement $\log M$,
     as a function of $\kappa$ for three values of the pairing
     interaction strengths $\eta=g\sqrt{\Omega}/\chi$ indicated in the insert.}
    \label{gsfig8}
  \end{minipage}
\end{figure}

The shift in $\omega_{\rm res}(g)$  due to the pairing interaction
is further illustrated in Fig. \ref{gsfig7}, which shows the
magnitude of the joint coherence function, $|G_{am}|$, as a
function of $\kappa=\omega/\chi\sqrt{\Omega}$ for three values of
the pairing coefficient $g$. Finally, Fig. \ref{gsfig8} shows the
entanglement entropy $E(\rho_b)$ of the ground state, obtained
from the von Neumann entropy of the molecular reduced density
operator~\cite{Hines}
\[
\rho_b=Tr_{f}(\rho),
\]
as
\begin{equation}
E(\rho_b)=-\sum_{n_b=0}^{M} |C(n_b)|^2 \log{|C(n_b)|^2},
\end{equation}
where the logarithm is taken in base $2$, for three values of the
pairing interaction strengths, $\eta=0, 5, 10$. The entanglement
in Figure~ \ref{gsfig8} is divided by a maximum entanglement,
$\log{M}$. Consistently with the results of of Figs. \ref{gsfig5}
and \ref{gsfig6} for the molecular statistics, the pairing
interaction reduces the entanglement.

\section{Dynamics}

In the absence of  pairing interaction our model is equivalent to
the Tavis-Cummings model, whose dynamics has been studied in
detail in the context of coherent spontaneous emission from a
system of $N$ two-level atoms interacting with a quantized
radiation field~\cite{Dicke,Bonifacio}. The main purpose of this
section is to extend this work to the study of the system dynamics
in the presence of $V_p$. One important result is that the
nonlinearity of this interaction produces a self-trapping
transition.

We assume that the system consists initially either of atomic
pairs or of molecules only, corresponding to a maximal spin state.
A photoassociation beam is applied from $t=0$ on, and we examine
the subsequent coherent dynamics of the system. Note that although
this problem resembles the dynamics of a bosonic Josephson
Junction ~\cite{BJJ1,BJJ2} in an asymmetric trap for the initial
imbalance in populations, the nonlinear coupling term in our model
leads to considerably richer dynamics. Since the total spin is a
constant of motion of the Hamiltonian~(\ref{DGmodel}), we confine
our discussion to the maximal spin state $|S=\Omega/2\ket$.

\subsection{Semiclassical approximation}

Before proceeding with a full quantum analysis, we first consider
the semiclassical approximation. Our approach is very similar to
that taken in Ref.~\cite{AMBEC2}. Introducing the two operators
\begin{eqnarray}
J_+&\equiv& S^+b,\nonumber \\
 J_-&\equiv&  b^\dagger S^-,
\end{eqnarray}
results in the Heisenberg equations of motion
\begin{eqnarray}
\label{heisenbergeq}
i\frac{d}{dt}J_{+}&=&\omega J_{+}
 +2\chi b^\dagger b S^z+\chi S^+ S^- -2gJ_{+}S^z , \\
i\frac{d}{dt}J_{-}&=&-\omega J_{-}
-2\chi b^\dagger b S^z
-\chi S^+ S^- +2g S^zJ_-, \\
i\frac{d}{dt}S^z&=&\chi (J_+ - J_-).
\end{eqnarray}
Here we note the following relations between operators and conserved
quantities,
\[
S^+ S^- = S(S+1)-S^z(S^z-1), \quad b^\dagger b = M-S-S^z.
\]

As usual we introduce a semiclassical approximation by factorizing
the mean values of the various operators that appear in the
Heisenberg equations of motion, such as $\bra S_z J_+\ket=\bra S^z
\ket\bra J_+ \ket$ and $\bra J_- S_z\ket=\bra J_-\ket\bra S_z
\ket$, etc. Introducing the $c-$number functions $s_z\equiv\bra
S_z\ket$, $j_x\equiv \bra (J_{+} + J_{-})\ket/2$, and $j_y\equiv
\bra (J_{+} - J_{-})\ket/2i$, and neglecting corrections of order
$1/\Omega$, that is, setting $\bra S^{+}S^{-}\ket=S^2-s_z^2$,
we obtain the semiclassical equations of motion
\begin{eqnarray}
\label{cdynamics}
\frac{d}{dt}j_x&=&\omega j_y-2gj_ys_z\\
\frac{d}{dt}j_y&=&-\omega j_x+2gj_xs_z-\chi h(s_z)\\
\frac{d}{dt}s_z&=&2\chi j_y,
\end{eqnarray}
where $h(s_z)=-3s_z^2+2(M-S)s_z+S^2$. Noting the additional
conserved quantity
\[
\frac{d}{dt}\left(j_x-\frac{\omega}{2\chi}s_z+\frac{g}{2\chi}s_z^2\right)=0,
\]
we find that the coupled equations~(\ref{cdynamics}) are
equivalent to the classical Newtonian equation of motion
\begin{equation}
\label{Newtoneq} \frac{d^2}{dt^2}s_z=-\frac{d}{ds_z}U(s_z),
\end{equation}
where the potential $U(s_z)$ is determined by the initial conditions
for $j_x(0)$ and $s_0\equiv s_z(0)$. It is sufficient for our
purpose to assume an initial Fock state, so that $j_x(0)=0$ and
$U(s_z)$ is given by
\begin{eqnarray}
\label{scdpotential}
U(s_z)&=&\frac{g^2}{2}s_z^4-(g\omega+2\chi^2)s_z^3\nonumber\\
&+&\left[\frac{\omega^2}{2}+g\omega
s_0-g^2(s_0)^2+2\chi^2(M-S)\right]s_z^2 \nonumber \\
&+&\left[2\chi^2 S^2-\omega^2s_0+\omega gs_0^2 \right]s_z.
\end{eqnarray}
Since the potential has a quartic form of $s_z$,
Eq.~(\ref{Newtoneq}) can be solved analytically in terms of
Jacobian elliptic functions. The derivation of the general
solutions, which is straightforward but lengthy, is given in
Appendix~A.

\subsection{Coherent dynamics, $g=0$}

We first examine a typical behavior of the semiclassical dynamics
in the absence of pairing interaction, $g=0$, for $M/\Omega\ll 1$,
$M/\Omega \sim 0.5$, and $M/\Omega\simeq 1$.

From the semiclassical solutions~(\ref{apg0eq3})
in the case of $g=0$ and on the exact resonance $\omega=0$,
the population imbalance between fermionic pairs and molecules,
$\bra \hat{n}_p\ket - \bra \hat{n}_b \ket=2S-M+2s_z$, and the coherence
function $j_y$ is given by
\begin{eqnarray}
\label{scdresonance1}
\frac{\bra n_p\ket-\bra n_b\ket}{M}&=&2\,{\rm sn}^2(\chi\sqrt{\Omega}t+\phi;k)-1,\\
\label{scdresonance2}
\frac{2j_y}{M\sqrt{\Omega}}&=&2\,{\rm sn}(\chi\sqrt{\Omega}t+\phi;k)
\times{\rm cn}(\chi\sqrt{\Omega}t+\phi;k)\nonumber\\
&\times&{\rm dn}(\chi\sqrt{\Omega}t+\phi;k),
\end{eqnarray}
where {\rm sn}, {\rm cn}, and {\rm dn} are Jacobian elliptic
functions~\cite{elliptic}. Here, $k=\sqrt{M/\Omega}$ is the
elliptic modulus and $\phi$ is a phase factor determined by the
initial conditions. It is equal to $\phi=K$, corresponding to the
complete elliptic integral of the first kind, for an initial
fermionic Fock state $|n_p=M\ket$, and to $\phi=0$ for an initial
the entire population being in the molecular mode. We note that
$j_x(t)$ is zero for all time on the exact resonance.

\subsubsection{$M/\Omega \ll 1$ --- Linear coupled-bosons regime}

In this case, the elliptic modulus $k\ll 1$ (see Appendix~A) so that
the elliptic functions in Eqs.~(\ref{scdresonance1}) and (\ref{scdresonance2})
can be approximated by ${\rm sn}(u,k)\to \sin{u}$, ${\rm cn}(u,k)\to
\cos{u}$, and ${\rm dn}(u,k)\to 1$, respectively. The imbalance
in atomic and molecular populations which undergoes Rabi
oscillations at the frequency $2\chi\sqrt{\Omega}$ is given by
    \[
        (\bra\hat{n}_p\ket-\bra\hat{n}_b\ket)/M=\cos{(2\chi\sqrt{\Omega}t)}
    \]
for an initial fermionic pair state,
    \[
    (\bra\hat{n}_p\ket-\bra\hat{n}_b\ket)/M=-\cos(2\chi \sqrt{\Omega}t)
    \]
for an initial molecular state. These results are equivalent to
those obtained directly from the Hamiltonian~(\ref{linearboson}).
We have compared these solutions with the exact quantum mechanical
dynamics obtained numerically, and checked that the linear
approximation agrees with the quantum results for
$M/\Omega\lesssim 0.2$ and for times shorter than $t\sim
\pi/\chi\sqrt{\Omega}$.

\subsubsection{$M/\Omega \sim 0.5$ --- Intermediate regime}

\begin{figure}
    \begin{center}
      \includegraphics[width=7.5cm,clip]{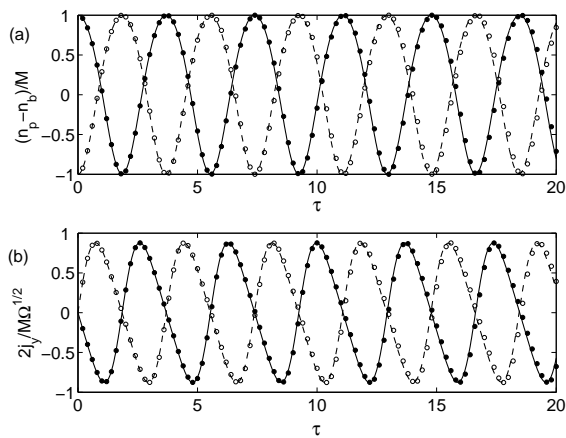}
    \end{center}
    \caption{Comparison of the semiclassical and quantum dynamics
        for $2M=\Omega=120$, $\omega=0$, and $g=0$.
        Figure (a) shows the population difference between fermionic pairs and
        molecules $(\bra\hat{n}_p\ket-\bra\hat{n}_b\ket)/M$. Figure (b) plots
        the coherence function $2j_y/M\sqrt{\Omega}$
        as a function of the dimensionless time $\tau=\chi\sqrt{\Omega}t$.
        The solid and dashed lines give the quantum results for an
        initial fermionic pair sate and a pure molecular state, respectively.
        The corresponding semiclassical results are indicated by filled
        and open circles, respectively.}
    \label{dyfig1}
\end{figure}

Figure~\ref{dyfig1}a shows the normalized population difference
$(\bra\hat{n}_p\ket-\bra\hat{n}_b\ket)/M$, and Fig.~\ref{dyfig1}b
shows the normalized coherence function $2j_y/M\sqrt{\Omega/2}$,
as a function of the dimensionless time $\tau=\chi\sqrt{\Omega}t$
for a system initially either in a pure atomic state or a pure
molecular state and for $M/\Omega=0.5$. The circles correspond to
the semiclassical description, while the lines are the results of a
full quantum-mechanical analysis. The anharmonicity due to
the nonlinear atom-molecule coupling is clearly apparent, and also
shows that the semiclassical dynamics approximate the quantum
dynamics very well.

We note that the atomic pair state in the half-filled shell
corresponds to a Dicke superradiant state~\cite{Dicke}, which is
known from quantum optics to give rise to the strongest collective
enhancement of transition probabilities. This enhancement is
proportional to the product of the number of particle pairs $M$
and the number of hole pairs $\Omega-M$, and is maximum for
$M=\Omega/2$. From Eq.~(\ref{scdresonance1}), we have that for
sufficient short times $\tau=t/\chi\sqrt{\Omega}\ll 1$ the average
number of molecules builds up as
\begin{eqnarray*}
\bra n_b \ket&=&{\rm cn}^2(\chi\sqrt{\Omega}t+\phi)\\
&\rightarrow_{\tau\ll 1}& M\left(\Omega-M\right)(\chi t)^2
=\frac{\chi^2}{4}\Omega^2t^2.
\end{eqnarray*}

\subsubsection{$M/\Omega \simeq 1$ --- Binary atomic-molecular BEC}

As shown in Ref.~\cite{Vardi}, the coherent dynamics in this
regime is qualitatively very similar to that of binary condensate
of atoms and molecules. For $M =\Omega$, corresponding to $k=1$,
the population imbalance between atomic pairs and molecules is
given in the semiclassical approximation by
\begin{equation}
\frac{\bra n_p\ket-\bra
n_b\ket}{M}=2\,\tanh{^2(\chi\sqrt{\Omega}t+\phi)-1},
\end{equation}
indicating that the point $\bra n_p\ket=M$ is stationary. However,
the system is dynamically unstable against small fluctuations ,
see Ref.~\cite{AMBEC2} for a detailed discussion in the context of
binary condensates of atoms and molecules and Ref.~\cite{Walls} in
the context of second-harmonic generation.

\subsection{\label{selftrap}Self-trapping transition and quantum dynamics}

To conclude, we discuss the coherent dynamics of the system in the
presence of pairing interaction, $g\neq 0$, considering only
the case of half-filling for simplicity. We consider specifically
the example $2M=\Omega=120$, which corresponds to $n_F=14$, and
take an initial state as the Fock state $|n_p=M;n_b=0\ket$.
Figure~\ref{appfig1} of the appendix shows the phase diagram of
the semiclassical dynamics in the $\kappa-\eta$ space,
illustrating that a self-trapping transition~\cite{BJJ1,Scott}
takes place when varying the dimensionless detuning frequency
$\kappa=\omega/\chi\sqrt{\Omega}$, provided that the pairing
interaction strength $\eta=g\sqrt{\Omega}/\chi$ exceeds a critical
strength $\eta_c$. For the case of half-filling case, we find
$\eta_c=5.0302$.

\begin{figure}
    \begin{center}
      \includegraphics[height=4.5cm,width=7.5cm,clip]{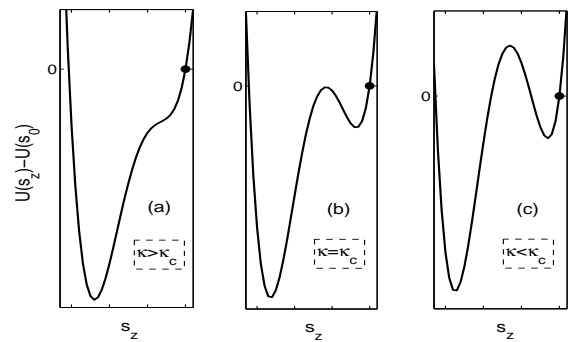}
    \end{center}
    \caption{Schematic potential curves, $U(s_z)-U(s_0)$, around
     a self-trapping transition point $\kappa_0$. These curves are
     at (a) above $\kappa_0$, (b) at transition point
     $\kappa=\kappa_0$, and (c) below $\kappa_0$. The filled circle
     indicates an initial position of classical particle.}
    \label{dyfig2}
\end{figure}
%\begin{widetext}
\begin{figure*}
    \begin{center}
      \includegraphics[height=12cm,width=18cm,clip]{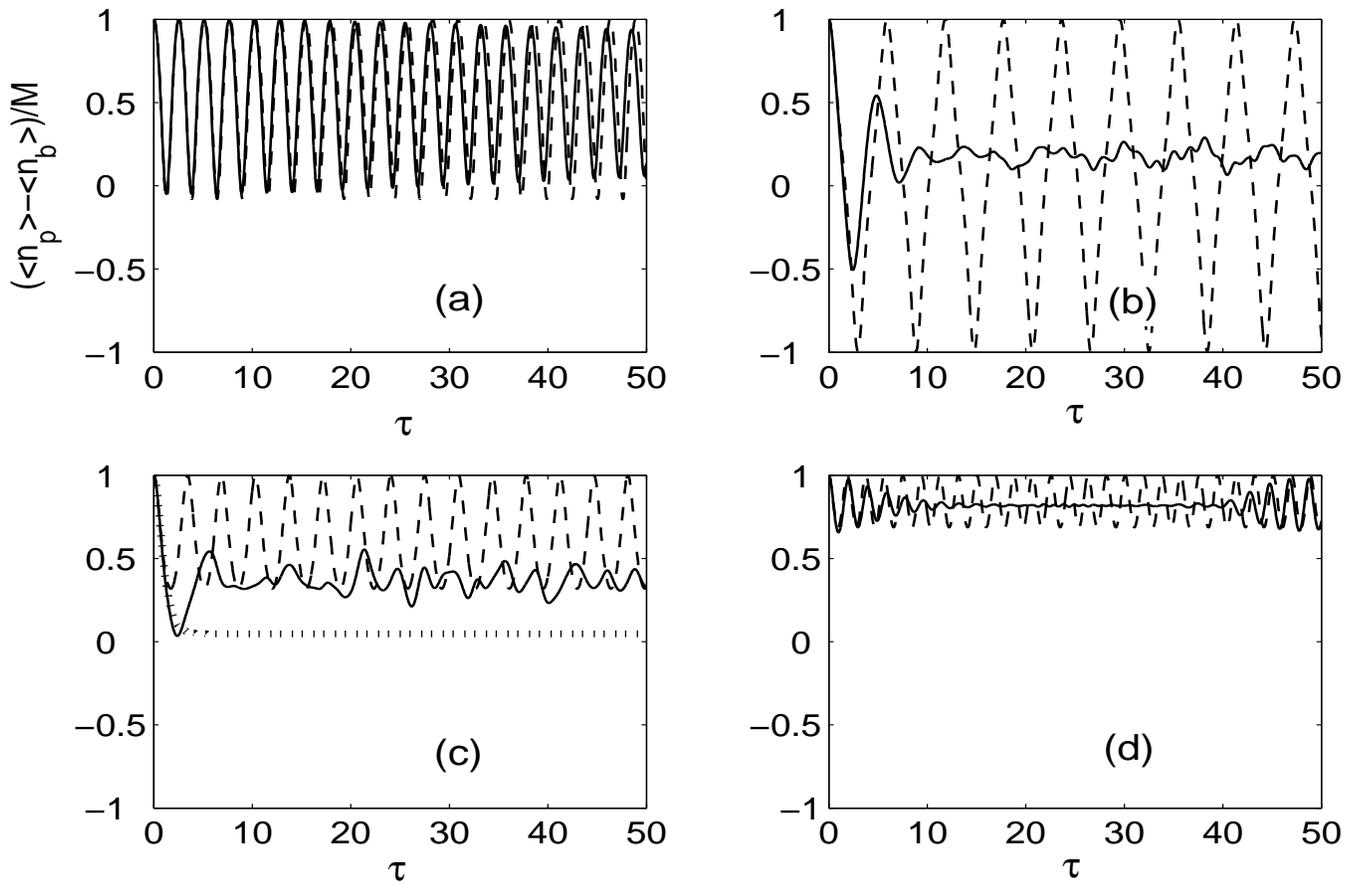}
    \end{center}
    \caption{Population imbalance versus the rescaled time
      $\tau=\chi\sqrt{\Omega}t$ for $\eta=6$. Quantum (solid line) and
      semiclassical
      (dashed line) solutions are shown, respectively, for the detuning
      parameters (a) $\kappa=0$ and (b) $\kappa=-3.1$, and
      (c) $\kappa=-3.3$, and (d) $\kappa=-4.0$.
      The dotted line in (c) corresponds to the semiclassical solution
      at the transition point $\kappa_0=-3.2307$.}
    \label{dyfig3}
\end{figure*}
\begin{figure*}
  \begin{minipage}{0.45\textwidth}
    \begin{center}
      \includegraphics[height=5cm,width=7.5cm,clip]{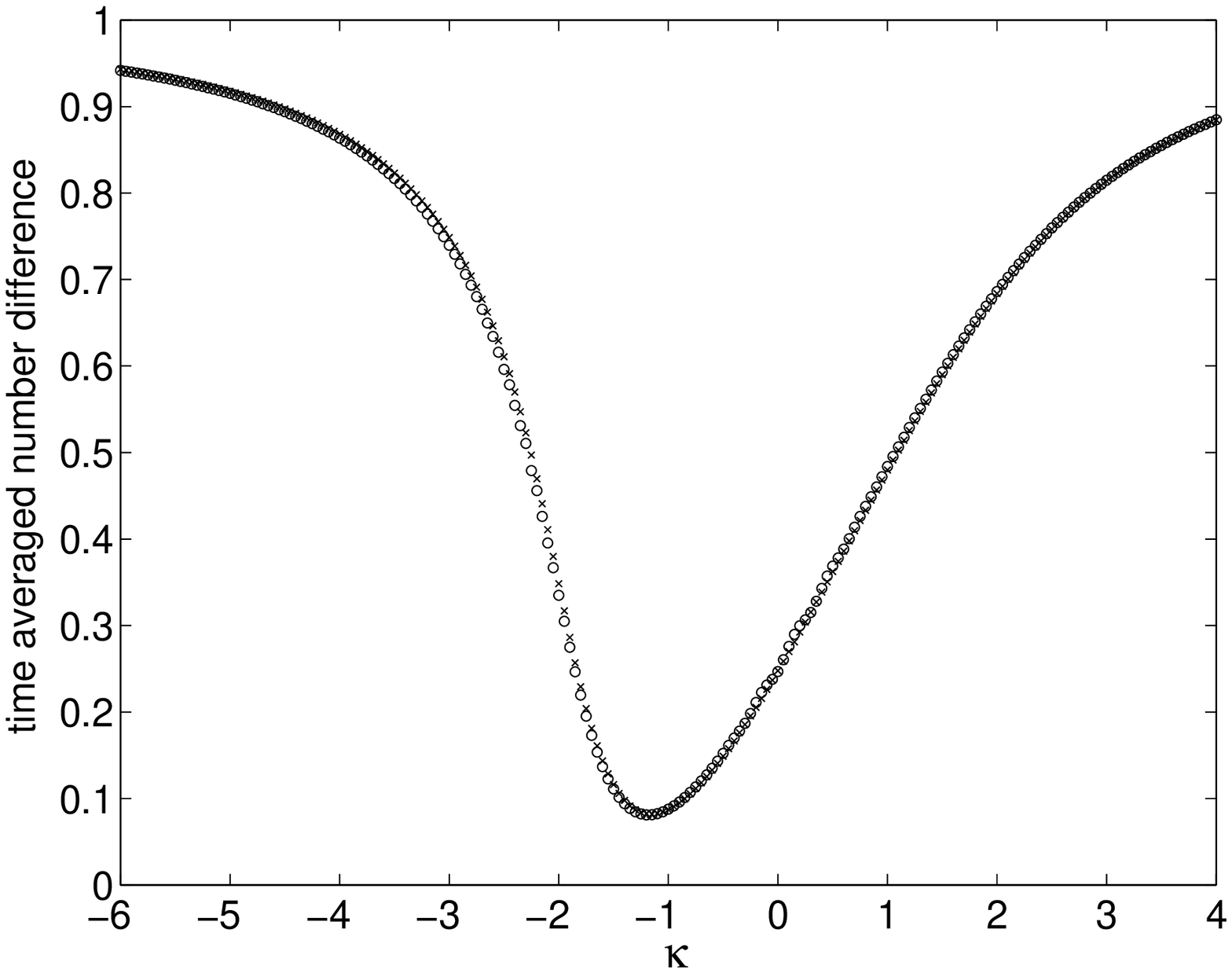}
    \end{center}
    \caption{Time-averaged population imbalance as a function of
      $\kappa$ for $\eta=2.0$.
    Quantum result (cross) and semiclassical result (open circle).}
    \label{dyfig4}
  \end{minipage}
  \hspace{4mm}
  \begin{minipage}{0.45\textwidth}
    \begin{center}
      \includegraphics[height=5cm,width=7.5cm,clip]{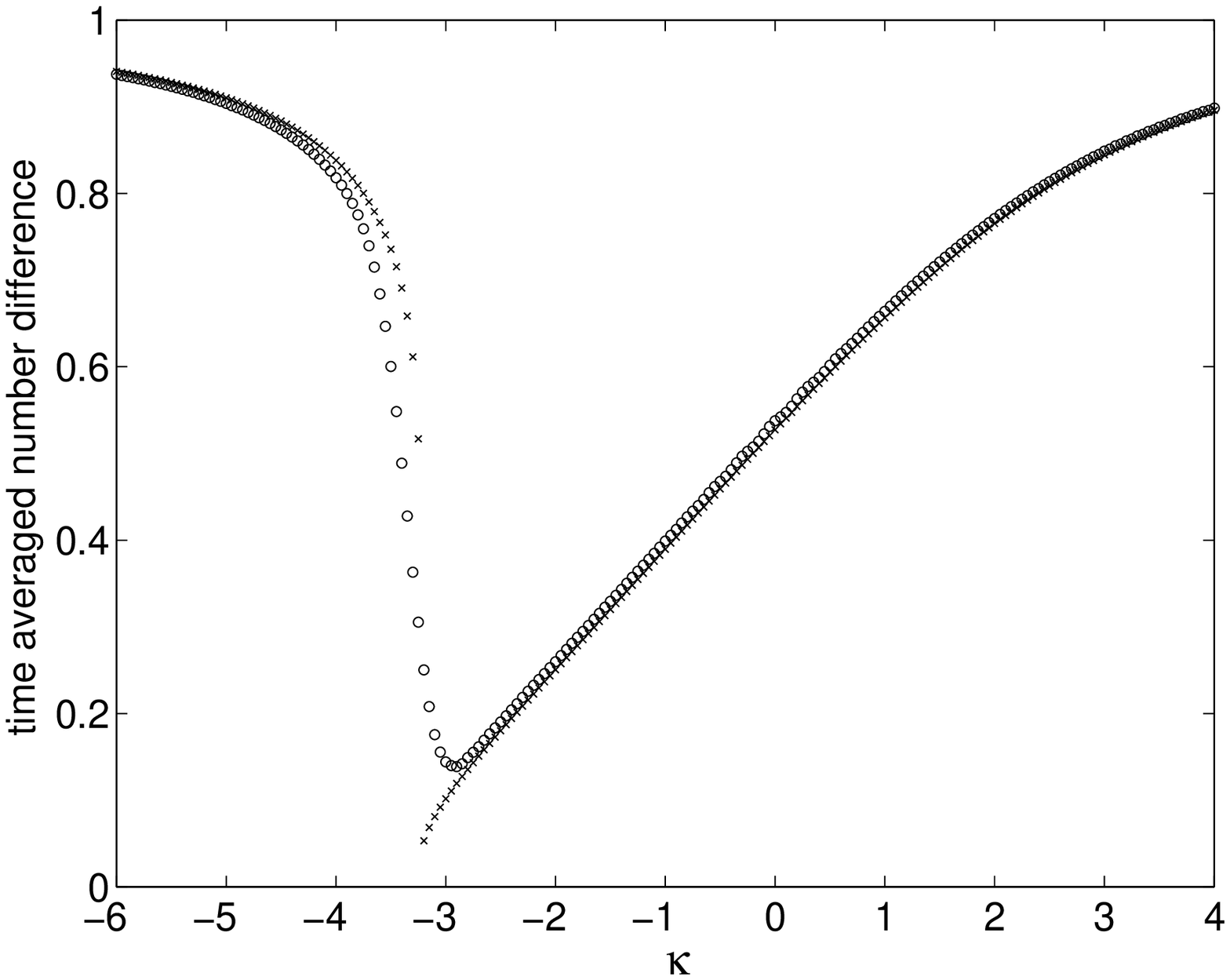}
    \end{center}
    \caption{Time-averaged population imbalance as a function of
      $\kappa$ for $\eta=6.0$.
    Quantum result (cross) and semiclassical result (open circle).}
    \label{dyfig5}
  \end{minipage}
\end{figure*}

This transition can be interpreted physically from the motion of
``classical particle''  in the ``potential'' (\ref{scdpotential}).
Figure~\ref{dyfig2} displays schematic potential curves,
$U(s_z)-U(s_0)$ as a function of $s_z$ in the vicinity of a
self-trapping transition point $\kappa_0$. For our specific
initial conditions the particle ``velocity'' is initially zero,
$ds_z(0)/dt\propto j_y(0)=0$. For $\kappa > \kappa_0$,
Fig.~\ref{dyfig2}-(a), the classical particle oscillates
periodically in the potential. As $\kappa$ approaches $\kappa_0$,
one additional potential barrier appears, and at the transition
point its height equals the initial potential energy of the
particle (see Fig.~\ref{dyfig2}-(b)). At that point, the particle
rests on the potential maximum after reaching it.  Below the
critical point, the barrier confines the particle in a narrow
range, as shown in Fig.~\ref{dyfig2}-(c). The self-trapping effect
provides a sudden suppression of the amplitude of coherent
oscillations. Since the key factor in achieving this transition is
the quartic term in the potential~(\ref{scdpotential}), it
disappears in the absence of pairing interaction.

Fig.~\ref{dyfig3} shows the time evolution of the population
difference $(\bra \hat{n}_{p}\ket-\bra \hat{n}_b\ket)/M$ for
$\eta=6$ and for several detuning energies. The semiclassical
solutions (dashed lines) clearly show the self-trapping as a
sudden suppression of coherent oscillations from just above to
just below the transition detuning. The dotted line in
Fig.~\ref{dyfig3}-(c) corresponds to the semiclassical solution
for the threshold detuning $\kappa_0=-3.2307$. The solid lines
show the exact quantum solutions. Apart from the transition point,
the quantum and semiclassical dynamics are similar at least for
short enough times. However, the oscillations of the quantum
solution deviate from those of the semiclassical solution near the
transition point, Fig.~\ref{dyfig3}(b-c). Since in the semiclassical
picture the height of the potential barrier near the transition
point is just below or above the initial potential energy of a
particle, the quantum motion of the particle is very sensitive to
fluctuations and hence deviates significantly from its classical
counterpart. The initial Fock state provides fluctuations of
coherence, and large quantum fluctuations of the population
imbalance arise as a result. We have verified numerically that the
number fluctuations near the transition point are enhanced by an
order of magnitudes as compared to those far away from that point.

Figs.~\ref{dyfig4} and \ref{dyfig5} compare the quantum and
semiclassical time-averaged population imbalance as a function of
$\kappa$ for $\eta=2.0$, and for $\eta=6.0$, respectively. In contrast to the
second case, there is no self-trapping transition in the first
case. Hence the semiclassical time-averaged population imbalance
(cross) is a smooth function of $\kappa$, and agrees well
with the quantum results. For the strong pairing coupling of
Fig.~\ref{dyfig5}, in contrast, an abrupt jump of the
semiclassical time-averaged value occurs when varying $\kappa$, a
signature of the self-trapping transition. Due to the large
quantum fluctuations, it differs markedly from the time-averaged
quantum result near a transition point.

%
%\begin{figure}
%  \begin{minipage}{0.45\textwidth}
%    \begin{center}
%      \includegraphics[width=7.5cm,clip]{dyfig2a.eps}
%    \end{center}
%    \caption{Dynamics for $g/\chi=1$ and for $\omega/\chi\Omega=0$.}
%    \label{dyfig2a}
%  \end{minipage}
%  \hspace{4mm}
%
%  \begin{minipage}{0.45\textwidth}
%    \begin{center}
%      \includegraphics[width=7.5cm,clip]{dyfig2b.eps}
%    \end{center}
%    \caption{Dynamics for $g/\chi=1$ and for $\omega/\chi\Omega=-0.4$}
%    \label{dyfig2b}
%  \end{minipage}
%\end{figure}
%
%\begin{figure}
%  \begin{minipage}{0.45\textwidth}
%    \begin{center}
%      \includegraphics[width=7.5cm,clip]{dyfig2c.eps}
%    \end{center}
%    \caption{Dynamics for $g/\chi=1$ and for $\omega/\chi\Omega=-0.47$.}
%    \label{dyfig2c}
%  \end{minipage}
%  \hspace{4mm}
%
%  \begin{minipage}{0.45\textwidth}
%    \begin{center}
%      \includegraphics[width=7.5cm,clip]{dyfig2d.eps}
%    \end{center}
%    \caption{Dynamics for $g/\chi=1$ and for $\omega/\chi\Omega=-0.6$}
%    \label{dyfig2d}
%  \end{minipage}
%\end{figure}

\section{Summary}

In this paper, we have considered the coherent photoassociation of
fermionic atoms into bosonic molecules trapped in a spherically
symmetric harmonic trap. We showed that under a realistic set of
conditions this system can be mapped onto a Tavis-Cummings
Hamiltonian with an additional paring interaction using
pseudo-spin operators. We carried out an exact numerical
diagonalization of the Hamiltonian to determine the ground state
of the system, investigating the crossover from a predominantly
atomic to a predominantly molecular state. We also investigated
the joint coherence and the quantum entanglement between the
atomic and molecular fields, and found that the atomic pairing
interaction suppresses the entanglement between fermions and
bosons. We then analyzed the coherent dynamics of photoassociation
due to the nonlinear atom-molecule coupling. Using a semiclassical
factorization ansatz, we showed the appearance of a self-trapping
transition in the presence of pairing interaction. An exact
quantum solution illustrated the important role of quantum
fluctuations in the neighborhood of that transition point. Future
work will extend this study to a detailed analysis of the
non-degenerate model and to multi-well superradiant systems. For
instance, preparing an atomic Fermi gas in a Josephson-type
configuration and applying a photoassociation beam should lead to
the efficient production of spatially correlated molecules.

\acknowledgements This work is supported in part by the US Office
of Naval Research, the National science Foundation, the US Army
Research Office, NASA, and the Joint Services Optics Program.

\appendix
\section{Analytic solutions in terms of Jacobian Elliptic
functions}

In this appendix, we obtain analytic solutions of semiclassical
dynamics that obey the Newtonian equation of motion
Eq.~(\ref{Newtoneq}), and show the phase diagram of
the semiclassical dynamics based on those solutions.

The solution of Eq.~(\ref{Newtoneq}) has the general form
\begin{equation}
\label{integral}
t=\int^{s_z}_{s_0}\frac{ds}{\sqrt{2[U(s_0)-U(s_z)}]},
\end{equation}
with the potential $U(s_z)$ given by Eq.~(\ref{scdpotential}).

We analyze the solution for the two cases $g=0$ and $g \neq 0$,
separately, and for simplicity we take the initial state as an
atomic state in the maximum spin manifold $S=\Omega/2$. The
extension to other initial states is straightforward.

\subsection{Case $g=0$}
In this case, the ``potential'' $U(s_z)$ is a cubic function of
$s_z$,
\begin{equation} U(s_z)=-2\chi^2
s_z^3+\left[\frac{\omega^2}{2}+2\chi^2 s_0\right]s_z^2
+\left[2\chi^2S^2-\omega^2s_0\right]s_z.
\end{equation}
Here we have used the initial condition $s_0=-S+M$. By introducing
the normalized quantities
   \[
      \bs_z=s_z/2S\quad (-1/2\leq \bs_z\leq 1/2), \qquad
      \bs_0=s_0/2S,
   \]
we obtain the explicit form of the denominator of the right-hand
side of Eq. (\ref{integral}),
\begin{eqnarray}
\varphi(s_z)&\equiv&2[U(s_0)-U(s_z)]\nonumber\\
&=&4\chi^2(2S)^3(\bs_z-\bs_0)(\bs_z-\bs_+)(\bs_z-\bs_-)\\
\bs_{\pm}&=&\frac{\kappa^2}{8}\pm\frac{1}{8}
\sqrt{\kappa^4-16\bs_0\kappa^2+16},
\end{eqnarray}
where $\kappa=\omega/\chi\sqrt{2S}$. We note that the variables
$\bs_\pm$ are always real-valued for any magnitudes of $\kappa^2$,
because $-1/2\le \bs_0 \le 1/2$, and then $\bs_+\ge \bs_0\ge
\bs_z(t) \ge \bs_-$. The solution can be obtained by integrating
the form
\begin{widetext}
\begin{equation}
\label{apg0eq0}
t= \int^{s_z}_{s_0}\frac{ds}{\sqrt{\varphi(s_z)}}
=\frac{1}{\chi\sqrt{2S(\bs_{+} -\bs_{-})}}
\left\{\int^{\phi}_{0}-\int^{\pi/2}_{0} \right\}
\frac{d\theta}{\sqrt{1-k^2 \sin^2{\theta}}},
\end{equation}
\end{widetext}
where,
\[
  \phi(t)=\arcsin{\sqrt{\frac{(\bs_+-\bs_-)(\bs_0-\bs_z)}{(\bs_0-\bs_-)(\bs_+-\bs_z)}}}.
\]
The integral that appears in that equation is an elliptic integral
of the first kind. Noting that the integration within $0\le
\theta\le \pi/2$ gives rise to a complete elliptic integral of the
first kind, $K$, we find
\begin{equation}
\chi\sqrt{2S(\bs_+ -\bs_-)} t+K={\rm sn}^{-1}\left(\sin{\phi(t)};k\right)
\end{equation}
where the function ${\rm sn}^{-1}$ is the inverse of the Jacobian
elliptic function and $k=\sqrt{(\bs_0-\bs_-)/(\bs_+-\bs_-)}$
denotes the elliptic modulus. This gives the evolution of
$\bs_z(t)$
\begin{equation}
    \label{apg0eq1}
    \bs_z(t)=\bs_- +
    (\bs_0-\bs_-){\rm sn}^2\left(\chi\sqrt{2S(\bs_+ - \bs_-)}t+K;k
    \right).
\end{equation}
At the exact resonance, $\omega=0$, we have that  $s_+=1/2$,
$s_-=-1/2$, and $k=\sqrt{M/2S}$ so that this expression reduces to
\begin{equation}
\label{apg0eq3} s_z(t)=-S+M{\rm sn}^2(\chi\sqrt{2S}t+K;k).
\end{equation}
In terms of $s_z(t)$ the coherence functions $j_x(t)$ and $j_y(t)$
are given by
\begin{equation}
\label{apg0eq2}
j_x(t)=\frac{\omega}{2\chi}s_z(t), \quad j_y(t)=\frac{1}{2\chi}
\frac{ds_z(t)}{dt}.
\end{equation}
This results in the expressions for the difference in atomic and
molecular populations $\bra n_{p} \ket-\bra n_b\ket$, and the
coherence function $j_y$ of Eqs.~(\ref{scdresonance1}) and
(\ref{scdresonance2}), respectively.

\subsection{Case $g\neq 0$}
The presence of pairing interaction renders the potential
quartic in $s_z$, see Eq.~(\ref{scdpotential}) where we have again
assumed that the initial state is an atomic state of maximum spin,
$|S,-S+M\ket$. It is convenient to introduce the function
$f(\bs_z)$ as
\begin{equation}
\varphi(s_z)=-g^2(2S)^4(\bs_z-\bs_0)\cdot f(\bs_z),
\end{equation}
where
\begin{eqnarray}
f(\bs_z)&=&\bs_z^3+\alpha \bs_z^2+\beta \bs_z+\gamma \nonumber\\
&=&(\bs_z-\bs_1)(\bs_z-\bs_2)(\bs_z-\bs_3),
\end{eqnarray}
$\alpha=\bs_0-2\kappa/\eta-4/\eta^2$,
$\beta=-\bs_0^2+\kappa^2/\eta^2$, and
$\gamma=-\bs_0^3+2\kappa\bs_0^2/\eta-\kappa^2\bs_0/\eta^2+1/\eta$
with $\eta=g\sqrt{\Omega}/\chi$.

The variables $\bs_1$, $\bs_2$, and $\bs_3$, which correspond to
the roots of the cubic equation, $f(s_j)=0$, are obtained by
``Cardano's formula''. With $\xi\equiv e^{i2\pi/3}$,
$p=-\alpha^2/3+\beta$, and $q=2\alpha^3/27-\alpha\beta/3+\gamma$,
and also $D=-(4p^3+27q^2)$, those roots are given by
\begin{eqnarray}
s_j&=&-\frac{\alpha}{3}+\xi^{j-1}\left[-\frac{q}{2}+\frac{1}{6}\sqrt{-\frac{D}{3}}\right]^{1/3}\nonumber\\
&&\qquad+\xi^{1-j}\left[-\frac{q}{2}-\frac{1}{6}\sqrt{-\frac{D}{3}}\right]^{1/3},
\end{eqnarray}
where $j=1,2,3$.

The numbers of real and complex roots are determined by the sign
of the polynomial discriminant $D$. If (a) $D > 0$, all three
roots are real and unequal. If (b) $D<0$, one root is real and two
are complex conjugates. If (c) $D=0$, two roots are equal for
$q\neq 0$, and all roots are equal for $q=p=0$.

\subsubsection{Case $D > 0$}

Suppose that $\bs_0\ge \bs_z\ge \bs_a$, and $\bs_a>\bs_b>\bs_c$ or
$\bs_b>\bs_c>\bs_0$, where each $s_{a,b,c}$ corresponds to one of
the roots $s_j$'s. The solution of ~Eq.(\ref{integral}) reads then
\begin{widetext}
\begin{equation}
t=\int^{s_z}_{s_0}\frac{ds}{\sqrt{\varphi(s)}}
=-\frac{1}{gS\sqrt{(\bs_0-\bs_b)(\bs_a-\bs_c)}}\int^\phi_0
\frac{d\theta}{\sqrt{1-k^2\sin^2{\theta}}},
%={\rm sn}^{-1}\left(\sin{\phi(t)}\right),
\end{equation}
where
\begin{equation}
\phi(t)=\arcsin{\sqrt{\frac{(\bs_a-\bs_c)(\bs_0-\bs_z)}{(\bs_0-\bs_a)(\bs_z-\bs_c)}}},
\qquad
k=\sqrt{\frac{(\bs_0-\bs_a)(\bs_b-\bs_c)}{(\bs_0-\bs_b)(\bs_a-\bs_c)}},
\end{equation}
so that
%\begin{widetext}
\begin{equation}
\label{apgeq1}
\bs_z(t)=\frac{\bs_0(\bs_a-\bs_c)+\bs_c(\bs_0-\bs_a){\rm sn}^2\left(-gS\sqrt{(\bs_0-\bs_b)(\bs_a-\bs_c)}t;k\right)}
{(\bs_a-\bs_c)+(\bs_0-\bs_a){\rm sn}^2\left(-gS\sqrt{(\bs_0-\bs_b)(\bs_a-\bs_c)}t;k\right)}.
\end{equation}
\end{widetext}

\subsubsection{Case $ D<0$}

Letting $\bs_a$ label the real root and with
$\bs_b=\bs_c^*=\bs_R+i \bs_I$, we have that
$\varphi(s_z)=-g^2(2S)^4(\bs_z-\bs_0)(\bs_z-\bs_a)
\left\{(\bs_z-\bs_R)^2+\bs_I^2\right\}$. With the change of
variable from $\bs_z$ ($\bs_0\ge \bs_z \ge \bs_a$) to
\begin{equation}
\frac{\bs_0-\bs_z}{\bs_z-\bs_a}=\frac{A}{B}\frac{1-\cos{\phi}}{1+\cos{\phi}},
\end{equation}
where $A=\sqrt{(\bs_0-\bs_R)^2+\bs_I^2}>0$ and
$B=\sqrt{(\bs_a-\bs_R)^2+\bs_I^2}>0$, and taking the elliptic
modulus as
\begin{equation}
k=\sqrt{\frac{(\bs_0-\bs_a)^2-(A-B)^2}{4AB}},
\end{equation}
the integral Eq.~(\ref{integral}) can then be replaced by
\begin{widetext}
\begin{equation}
t=\int^{s_z}_{s_0}\frac{ds}{\sqrt{\varphi(s)}}
=-\frac{1}{2gS\sqrt{AB}}\int^{\phi}_0\frac{d\theta}{\sqrt{1-k^2\sin^2{\theta}}}.
%={\rm sn}^{-1}(\sin{\phi(t})),\\
\end{equation}
The semiclassical solution is then given by
\begin{equation}
\label{apgeq2}
\bs_z(t)=\frac{\bs_aA+\bs_0B-(\bs_aA-\bs_0B){\rm cn}\left(-2gS\sqrt{AB}t;k\right)}
{A+B-(A-B){\rm cn}\left(-2gS\sqrt{AB}t;k\right)}.
\end{equation}
\end{widetext}

\subsubsection{Case $ D=0$}

In this case, the solution of Eq.~(\ref{integral}) can be
expressed in terms of elementary functions. For $q\neq 0$ the
solutions are equivalent to Eq.~(\ref{apgeq1}) in which the
elliptic functions are replaced by trigonometric functions for
$k=0$ ($\bs_b=\bs_c$), or hyperbolic functions for $k=1$
($\bs_0>\bs_a=\bs_b$). If $q=p=0$, the function $f(\bs_z)$ has
triple degenerate roots at a point $\bs_a=\bs_b=\bs_c$, and the
corresponding solution is obtained by
\begin{equation}
\label{apgeq3}
\bs_z(t)=\bs_a+\frac{\bs_0-\bs_a}{1-\left\{-gS(\bs_0-\bs_a)t\right\}^2}.
\end{equation}

\subsection{\label{phasediagram}Phase diagram of semiclassical dynamics}

In this subsection, we discuss the structure of the semiclassical
dynamics in $\kappa-\eta$ parameter space for the specific case of
a half-filled shell, $2M=\Omega$, by calculating the elliptic
modulus of the semiclassical solution.

\begin{figure}
    \begin{center}
      \includegraphics[width=7.5cm,clip]{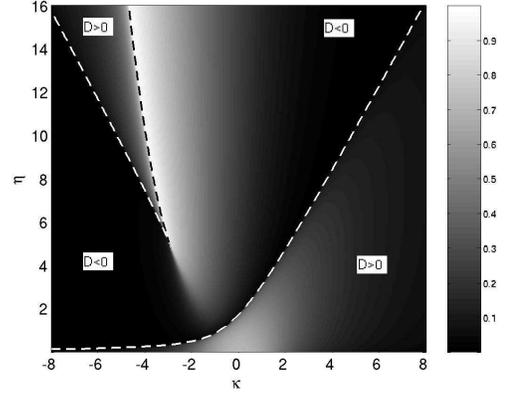}
    \end{center}
    \caption{Elliptic modulus in $\kappa-\eta$ space
      for the half-filled shell: $2M=\Omega$.}
    \label{appfig1}
\end{figure}

For $g=0$, Eq.~(\ref{apg0eq1}) describe all possible dynamics for
arbitrary detuning energy $\omega$, while in the presence of a
pairing interaction the dynamics is given by
solutions~(\ref{apgeq1}) and (\ref{apgeq2}), depending on the sign
of $D$. The dynamics is described by Eq.~(\ref{apgeq3}) at a
single singular point in $\kappa-\eta$ space, as we shall see
later.

Figure~\ref{appfig1} shows the elliptic modulus of the
semiclassical solution in $\kappa-\eta$ space. The two regions $D>
0$ and $D<0$ are separated by $D=0$ lines that correspond to the
specific values of elliptic modulus $k=0$ and $k=1$.
Eqs.~(\ref{apgeq1}) and (\ref{apgeq2}) coincide for $k=0$, which
corresponds to the white lines in Fig.~\ref{appfig1}. Hence these
two solutions connect continuously when crossing that line. For
the black line $k=1$, on the other hand, solutions corresponding
to Eq.~(\ref{apgeq1}) differ from Eq.~(\ref{apgeq2}). This
discontinuity gives rise to the self-trapping transition discussed
in subsection~\ref{selftrap}.

Figure~\ref{appfig1} shows that the black line and one of the
white lines intersect at the critical point $(\kappa_c, \eta_c)$,
where $q=p=0$, given explicitly by
\begin{eqnarray}
\eta_c&=&2\sqrt{\frac{2}{3}} (45+26\sqrt{3})^{1/4}\simeq 5.0302\\
\kappa_c&=&-\frac{4}{\eta_c}(2+\sqrt{3})\simeq -2.9677.
\end{eqnarray}
Form this result, we conclude that the self-trapping transition
appears by varying the detuning parameter $\kappa$ only for
$\eta>\eta_c$ .

\end{document}